\documentclass[letterpaper,twocolumn,10pt]{article}
\usepackage{usenix2019_v3}

\usepackage[normalem]{ulem}
\usepackage{amssymb}
\usepackage[english]{babel} 

\usepackage{xcolor} 
\usepackage{graphicx}
\usepackage{subfig}
\usepackage{multirow}
\usepackage{mathtools}
\usepackage{amsmath}
\usepackage{tabularx, booktabs}
\usepackage{colortbl}

\usepackage{algorithm}
\usepackage{algorithmic}

\DeclareMathAlphabet{\mathcal}{OMS}{cmsy}{m}{n}

\usepackage{cite}

\usepackage{xstring}
\usepackage{babel}
\usepackage{collcell}
\usepackage{hhline}
\usepackage{pgf}

\usepackage{balance}

\usepackage{caption}
\usepackage{nccmath}

\usepackage{xspace}
\newcommand{\Tool}{ALERT\xspace}
\newcommand{\ToolAlter}{ALERT*\xspace}
\newcommand{\ToolAny}{ALERT$_\text{Any}$\xspace}
\newcommand{\ToolTrad}{ALERT$_\text{Trad}$\xspace}

\begin{document}

\title{
\Tool: Accurate Learning for Energy and Timeliness
}

\author{
{\rm Chengcheng Wan, Muhammad Santriaji, Eri Rogers, Henry Hoffmann, Michael Maire, Shan Lu}\\
The University of Chicago
} 


\date{}
\maketitle

\begin{abstract}
 
An increasing number of software applications incorporate runtime Deep Neural Networks (DNNs) to process sensor data and return inference results to humans.  Effective deployment of DNNs in these interactive scenarios requires meeting latency and accuracy constraints while minimizing energy, a problem exacerbated by common system dynamics.

Prior approaches handle dynamics through either (1) system-oblivious DNN adaptation, which adjusts DNN latency/accuracy tradeoffs, or (2) application-oblivious system adaptation, which adjusts resources to change latency/energy tradeoffs. In contrast, this paper improves on the state-of-the-art by coordinating application- and system-level adaptation. \Tool, our runtime scheduler, uses a probabilistic model to detect environmental volatility and then simultaneously select both a DNN and a system resource configuration to meet latency, accuracy, and energy constraints.  We evaluate \Tool on CPU and GPU platforms for image and speech tasks in dynamic environments.  \Tool's holistic approach achieves more than 13\% energy reduction, and 27\% error reduction over prior approaches that adapt solely at the application or system level. Furthermore, \Tool incurs only 3\% more energy consumption and 2\% higher DNN-inference error than an oracle scheme with perfect application and system knowledge.

\end{abstract}


\section{Introduction}
\label{sec:intro}

\subsection{Motivation}

Deep neural networks (DNNs) have become a key workload  for many computing systems due to their high inference accuracy. This accuracy, however, comes at a cost of long latency, high energy usage, or both. Successful DNN deployment requires meeting a variety of user-defined, application-specific goals for latency, accuracy, and often energy in unpredictable,
dynamic environments.

Latency constraints naturally arise with DNN deployments when inference interacts with the real world as a consumer---processing data streamed from a sensor---or a producer---returning a series of answers to a human. For example, in motion tracking, a frame must be processed at camera speed \cite{jiang2018chameleon}; in simultaneous interpretation, translation must be provided every 2--4 seconds\cite{translation}. Violating these deadlines may lead to severe consequences: if a self-driving vehicle cannot act within a small time budget, life threatening accidents could follow \cite{lin2018architectural}.

Accuracy and energy requirements are also common and may vary for different applications in different operating environments. On one hand, low inference
accuracy can lead to software failures \cite{pei2017deepxplore, sun2018concolic}.
On the other hand, it is beneficial to minimize DNN energy or resource usage to extend mobile-battery time or reduce server-operation cost \cite{google-tpu}. 

These requirements are also highly dynamic. For example, the latency requirement for a
job could vary dynamically depending on how much time
has already been consumed by related jobs before it\cite{lin2018architectural}; 
the power budget and the accuracy requirement
for a job may switch among different settings
depending on what type of events are currently sensed \cite{apollo}. Additionally, the latency requirement may change based on the computing system's current context; e.g., in robotic vision systems the latency requirement can change based on the robot's latency and distance from perceived pedestrians \cite{dollar2011pedestrian}.

Satisfying all these requirements in a dynamic 
computing environment where the inference job may
compete for resources against unpredictable, co-located
jobs is challenging. Although prior work addresses these problems at either the application level or system level separately, each approach by itself lacks critical information that could be used to produce better results.

At the application level, different DNN designs---with
different depths, widths, and numeric precisions---provide various latency-accuracy trade-offs for the same inference task \cite{venkataramani2014axnn, judd2016proteus, hashemi2017understanding, jain2018compensated, sim2018dps}.
Even more dynamic schemes have been proposed
that adapt the DNN by dynamically changing its structure at the beginning of \cite{figurnov2017spatially, mcgill2017deciding, andreas2017convolutional, wu2018blockdrop} or during \cite{larsson2016fractalnet, branchynet, huang2017multi,lee2018anytime,wang2018anytime,hu2019learning,Our-icml-paper}  every inference tasks. 

Although helpful, these techniques are sub-optimal without considering system-level adaptation options.
For example, under energy pressure, these application-level adaptation techniques have to 
switch to lower-accuracy DNNs, sacrificing accuracy
for energy saving, even if the energy goal
could have been achieved by lowering the system power
setting (if there is sufficient latency budget).

At the system level, machine learning \cite{ansel2012siblingrivalry, lee2008efficiency, petrica2013flicker, ponomarev2001reducing, sridharan2013holistic,paragonASPLOS,quasarASPLOS,leoASPLOS2015}  and control theory  \cite{kalman2009,kalman2014, caloree, hoffmann2015, santriaji2016grape, zhou2016cash,poet,spectr} based techniques
have been proposed to dynamically assign system resources to better satisfy system and application constraints. 

Unfortunately, without considering the option of 
application adaptions, these techniques also reach sub-optimal solutions. For example, when the current DNN offers much higher accuracy than necessary, switching to a lower-precision
DNN may offer much more energy saving than any 
system-level adaptation techniques. This problem is exacerbated because, in the DNN design space, very small drops in accuracy enable dramatic reductions in latency, and therefore system resource requirements.

A cross-stack solution would enable
 DNN applications to meet multiple, dynamic constraints. 
However, offering such a holistic solution is non-trivial. 
The combination of DNN and system-resource adaptation creates a huge configuration space, making it difficult to dynamically and efficiently
predict which combination of DNN and system settings  will
meet all the requirements optimally.
Furthermore, without careful coordination, adaptations at the application and system level may conflict
and cause constraint violations, like missing a latency deadline
due to switching to higher-accuracy DNN
and lower power setting at the same time.

\subsection{Contributions}
This paper presents \Tool, a cross-stack runtime system for DNN inference to meet user goals by simultaneously adapting both DNN models and system-resource settings. 

\paragraph{Understanding the challenges}
We profile DNN inference across applications, inputs, hardware, and resource contention confirming there is a high variation in inference time.  This leads to
challenges in meeting not only 
latency but also energy and accuracy requirements.
Furthermore, our profiling of 42 existing DNNs for image classification confirms that different designs offer a wide spectrum of latency, energy, and accuracy tradeoffs. In general, higher accuracy comes at the cost of longer latency and/or higher energy consumption. These trade-offs offered
provide both opportunities
and challenges to holistic inference management
(Section \ref{sec:motivation}).

\begin{figure}
    \centering
    \includegraphics[width=0.95\linewidth]{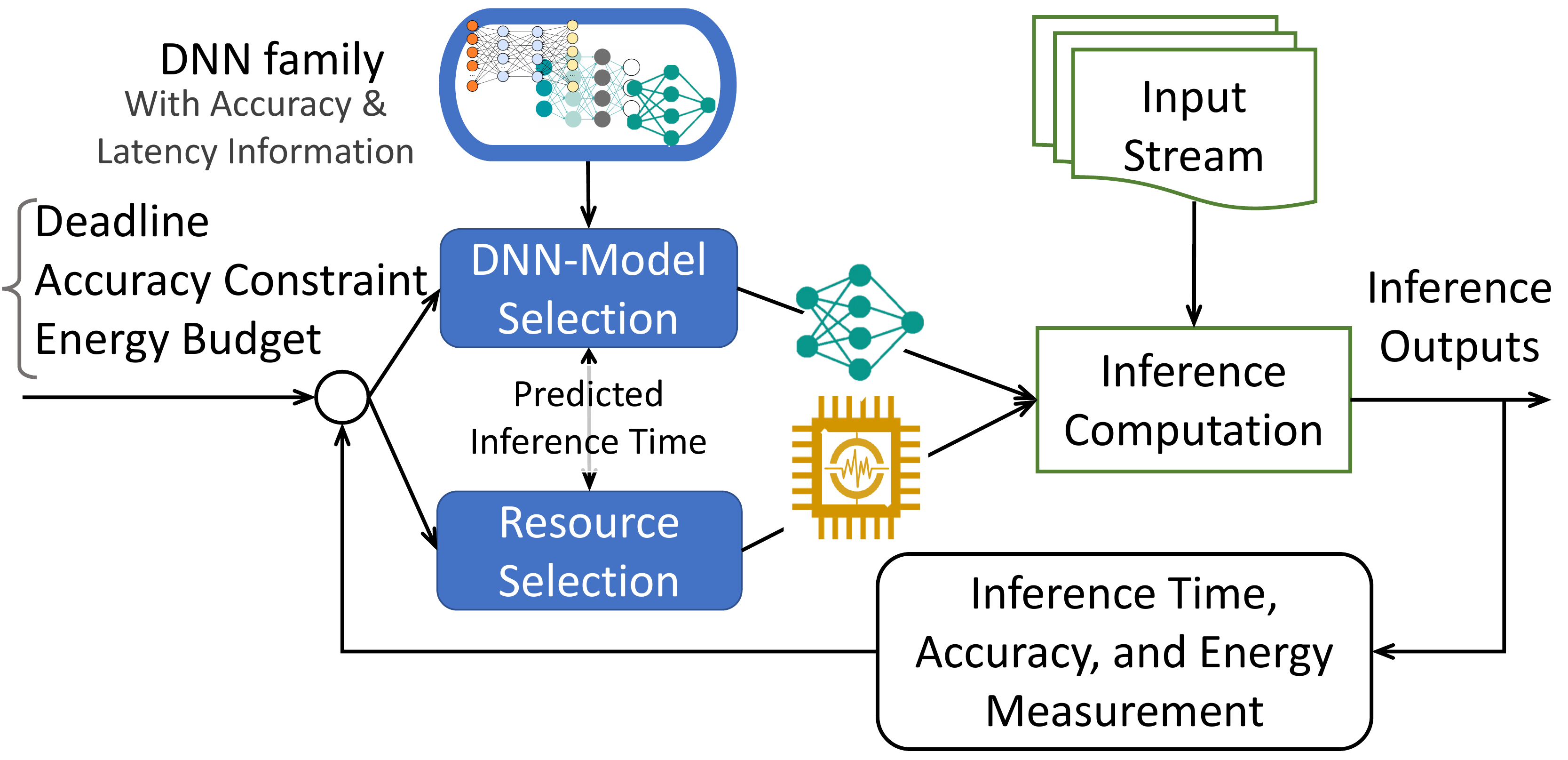}
    \vspace{-0.05in}
    \caption{\Tool inference system}
    \label{fig:architecture}
\end{figure}

\paragraph{Run-time inference management}
We design \Tool, a DNN inference management system that dynamically selects and adapts a DNN and a system-resource setting together to handle changing system environments and meet dynamic energy, latency, and accuracy requirements with probabilistic guarantees (Section \ref{sec:scheduler}).

\Tool is a feedback-based run-time. It measures inference accuracy, latency, and energy consumption; it checks whether the requirements on these goals are met; and, it then outputs both system and application-level configurations adjusted to the current requirements and operating conditions. \Tool focuses on meeting constraints\footnote{\Tool provides probabilistic, not hard guarantees, as the latter requires much more conservative configurations, often hurting both energy and accuracy. Section \ref{sec:design_limit} discusses this issue further.} in {\it any} two dimensions while optimizing the third; e.g., minimizing energy given accuracy and latency requirements or maximizing accuracy given latency and energy budgets. 

The key is estimating how DNN and system configurations interact to affect the goals.  To do so, \Tool addresses three primary challenges: (1) the combined DNN and system configuration space is huge, (2) the environment may change dynamically (including input, available resources, and even the required constraints), and (3) the predictions must be low overhead to have negligible impact on the inference itself.

\Tool addresses these challenges with a \emph{global slow-down factor}, a random variable relating the current runtime environment to a nominal profiling environment.  After each inference task, \Tool estimates the global slow-down factor using a Kalman filter.  The global slow-down factor's mean represents the expected change compared to the profile, while the variance represents the current volatility.  The mean provides a single scalar that modifies the predicted latency/accuracy/energy for \textit{every} DNN/system configuration---a simple mechanism that leverages commonality among DNN architectures to allow prediction for even rarely
used configurations (tackle challenge-1), while incorporating variance into predictions naturally makes \Tool conservative in volatile environments and aggressive in quiescent ones (tackle challenge-2).  
The global slow-down factor and Kalman filter are efficient to implement and low-overhead (tackle challenge-3).
Thus, \Tool combines the global slow-down factor with latency, power, and accuracy measurements to select the DNN and system configuration with the highest likelihood of meeting the constraints optimally.

We evaluate \Tool using various DNNs and application domains on different (CPU and GPU) machines under various constraints. 
Our evaluation shows that \Tool overcomes dynamic variability efficiently. 
Across various experimental settings, 
\Tool meets constraints in most cases while achieving within 93--99\% of optimal energy saving or accuracy optimization. 
Compared to approaches that adapt at application-level or system-level only \Tool achieves more than 13\% energy reduction, and 27\% error reduction 
(Section \ref{sec:experiment}).


\section{Understanding Deployment Challenges}

\label{sec:motivation}

We conduct an empirical study to 
examine the large
trade-off space offered by different DNN
designs and system settings (Sec. \ref{sec:back_tradeoff}), and the timing variability of 
inference 
(Sec. \ref{Sec:back_var}).

\begin{table}[htpb]
 \footnotesize  {
 \centering \addtolength{\tabcolsep}{-2pt}
\centering
\begin{tabular}{|l|c|c|c|c|}
\hline
       & Embedded                                                              & CPU1                                                       & CPU2                                                                     & GPU                                                        \\ \hline
CPU    & \begin{tabular}[c]{@{}c@{}}ARM \\Cortex  A-15\\ @2.0 GHz\end{tabular} & \begin{tabular}[c]{@{}c@{}}Core-i7\\ @2.2 GHz\end{tabular} & \begin{tabular}[c]{@{}c@{}}Xeon(R) \\ Gold 6126 \\ @2.60GHz\end{tabular} & \begin{tabular}[c]{@{}c@{}}Core-i7\\ @2.2 GHz\end{tabular} \\ \hline
GPU    & none                                                                  & none                                                       & none                                                                     & RTX 2080                                                   \\ \hline
Memory & DDR3 2G                                                               & DDR4 16G                                                   & DDR4 16G*12                                                              & DDR4 16G                                                   \\ \hline
LLC    & 2MB                                                                   & 9MB                                                        & 19.25MB                                                                  & 9MB                                                        \\ \hline
\end{tabular}
\vspace{-0.05in}
}
\caption{Hardware platforms used in our experiments} 
    \label{tbl:config}
    
\end{table}

\begin{table}[ht]
\footnotesize {
\centering
\begin{tabular}{|@{\hspace{0.3em}}c|l|l|l@{\hspace{0.3em}}|}
\hline
ID&Task                                                                                   & \multicolumn{1}{c|}{DNN Models}     & \multicolumn{1}{c|}{Datasets}                                                                                                    \\ \hline
IMG1&{Image}                                                   & VGG16 \cite{vgg} &ILSVRC2012\\ \cline{3-3}\cline{1-1}
IMG2& Classification & ResNet50 \cite{Resnet}       &  (ImageNet)       \\ 
\cline{2-4}\cline{1-1}
NLP1 & Sentence Prediction    & RNN                                    & Penn Treebank \cite{ptb}    \\ \hline
NLP2& Question & Bert \cite{bert}      &Stanford Q\&A\\ 
    & Answering & &  Dataset (SQuAD) \cite{rajpurkar2016squad}\\\hline
\end{tabular}
}
\vspace{-0.05in}
\caption{ML tasks and benchmark datasets in our experiments}
\label{tbl:task}
\end{table}

We use two canonical machine learning tasks, with state-of-the-art networks and common data-sets (see Table \ref{tbl:task}) on a diverse set of hardware platforms,
representing embedded systems, laptops (CPU1), CPU servers (CPU2), and GPU platforms
(see Table \ref{tbl:config}). 
The two tasks, image classification and natural language processing (NLP), are often deployed with deadlines---e.g., for motion tracking \cite{jiang2018chameleon} and simultaneous interpretation \cite{translation}---and both have received wide attention leading to a diverse set of DNN models.

\begin{figure}
    \centering
    \includegraphics[width=0.9\linewidth]{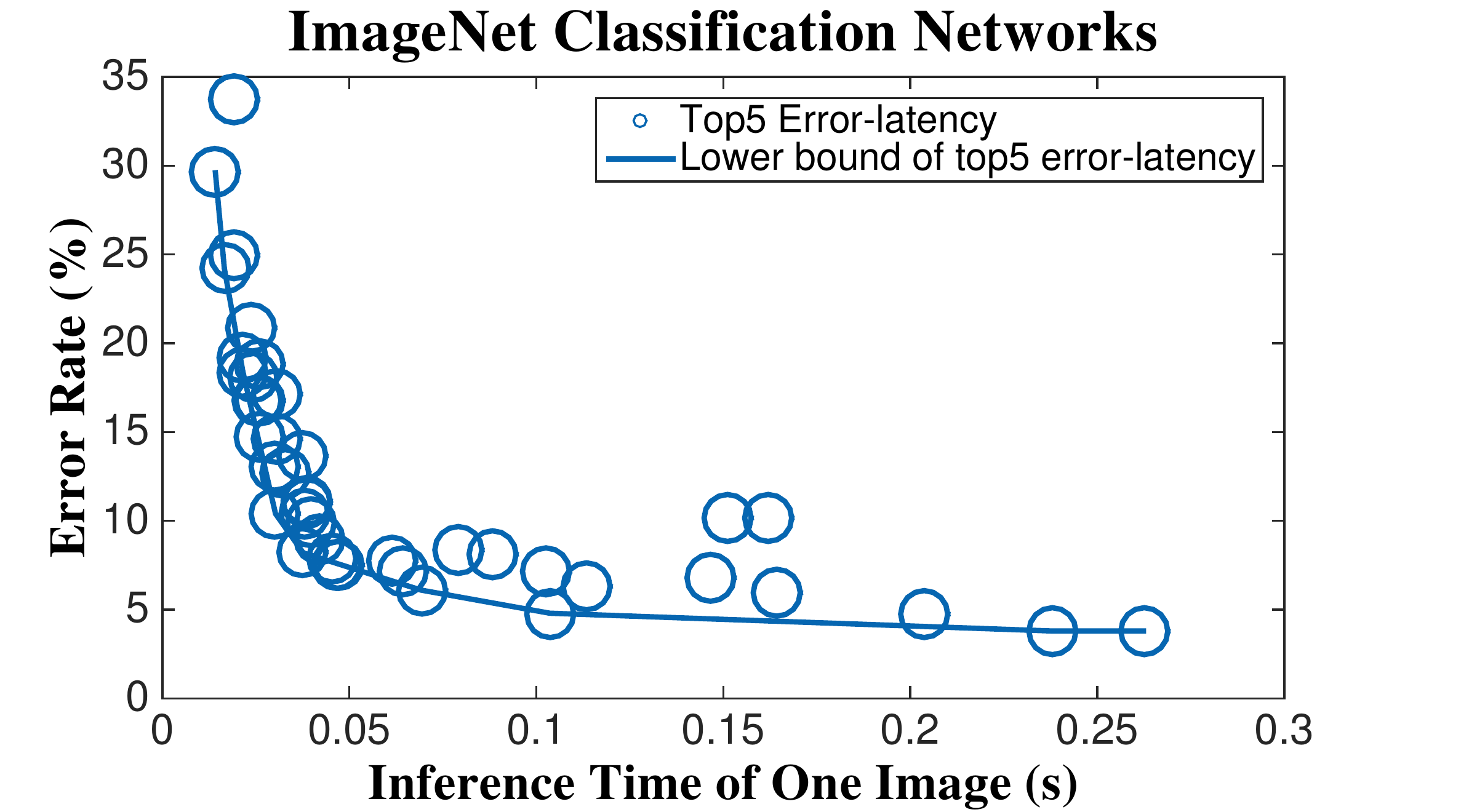}
    \vspace{-0.1in}
    \caption{Tradeoffs for 42 DNNs (CPU2). 
    }
    \label{fig:tradeoff_energy}
\end{figure}

\begin{figure}
    \centering
    \includegraphics[width=0.9\linewidth]{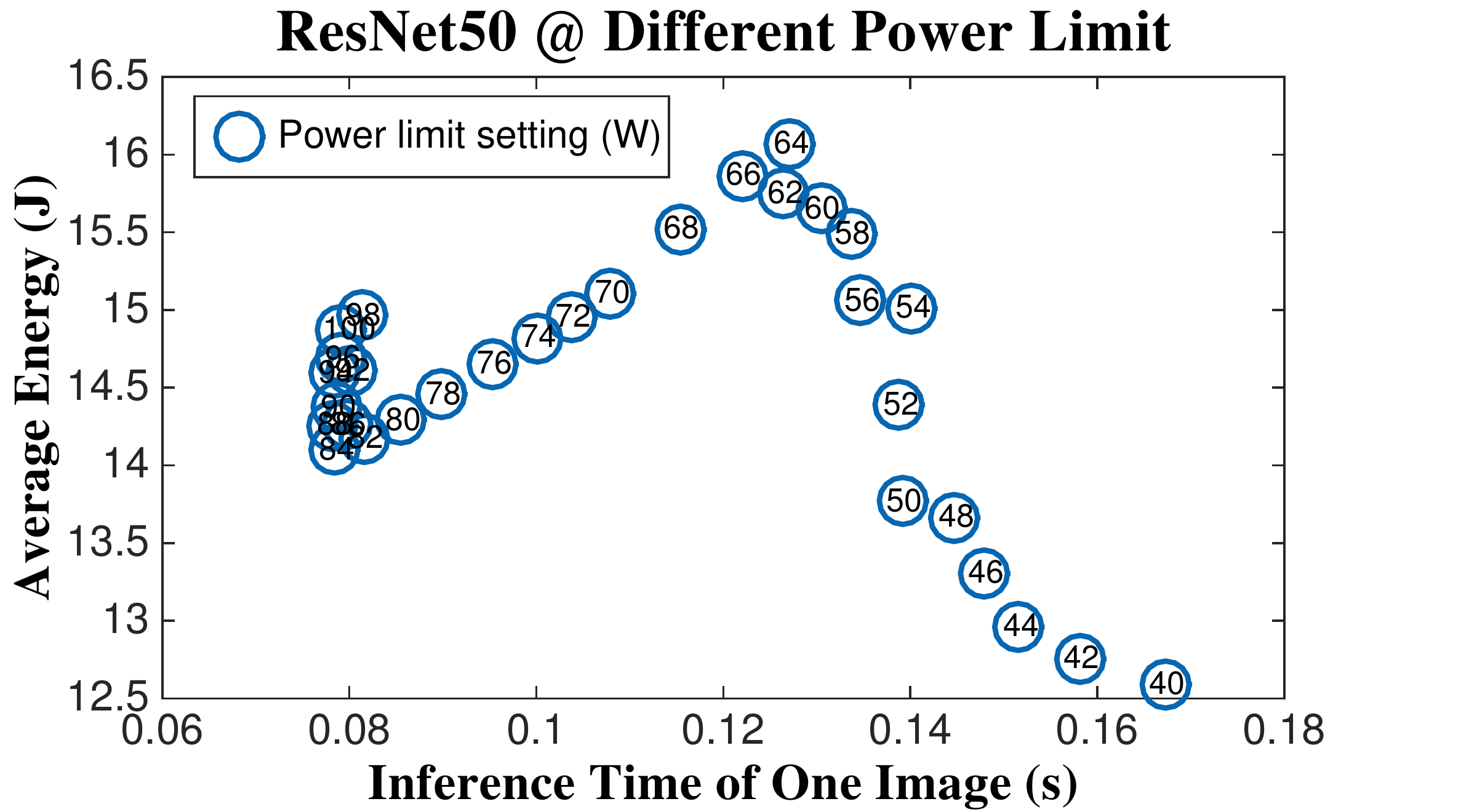}
    \vspace{-0.1in}
    \caption{Tradeoffs for ResNet50 at different power settings (CPU2). {(Numbers inside circles are power limit settings.)}}
    \label{fig:tradeoff_energy_latency}
\end{figure}

\subsection{Understanding the Tradeoffs}
\label{sec:back_tradeoff}

\textbf{Tradeoffs from DNNs}
We run \textbf{all} 42 image classification models provided by the Tensorflow website \cite{slim} on the 50000 images from ImageNet \cite{imangenet}, and 
measure their average latency, accuracy (error rate), and energy consumption. The results from CPU2
are shown in 
Figure \ref{fig:tradeoff_energy}. We can clearly see two trends from the figure, which
hold on other machines. 

First, different DNN models 
offer a \textit{wide} spectrum of accuracy
(error rate in figure), latency, and energy.
As shown in the figure,
the fastest model runs almost 18$\times$ faster than the slowest one and the most accurate
model has about 7.8$\times$ lower error rate than the least accurate.
These models also consume a wide range---more than $20\times$---of energy usage.

Second, there is no magic DNN that 
 offers both the best accuracy and the lowest latency, confirming the intuition that there exists a tradeoff between DNN accuracy and 
 resource usage. 
Of course, some DNNs offer better tradeoffs than others. 
In Figure \ref{fig:tradeoff_energy},  
all the networks sitting above the lower-convex-hull curve
represent sub-optimal tradeoffs.

\textbf{Tradeoffs from system settings}
We run ResNet50 under
31 power settings from 40--100W on CPU2.
We consider a sensor processing scenario with periodic inputs, setting the period to the latency under 40W cap.
We then plot the average energy consumed for the whole period (run-time plus idle energy) and the average inference latency
in Figure \ref{fig:tradeoff_energy_latency}. 

The results reflect two trends, which hold on other machines.
First, a large latency/energy space is available by changing system settings. The fastest setting (100W) is more than 2$\times$ faster than the slowest setting (40W). The most energy-hungry setting (64W) uses 1.3$\times$ more energy than the least (40W). 
Second, there is no easy way to choose the best setting. 
For example, 40W offers
the lowest energy, but highest latency. Furthermore, most of these points are sub-optimal in terms of energy and latency tradeoffs.  For example, 84W should be chosen for extremely low latency deadlines, but all other nearby points (from 52--100) will harm latency, energy or both.
Additionally, when deadlines change or when there is resource contention, the energy-latency curve also changes and different points become optimal.

\textbf{Summary:} DNN models and system-resource settings offer a huge trade-off space.
The energy/latency tradeoff space is not smooth (when accounting for deadlines and idle power) and optimal operating points cannot be found with simple gradient-based heuristics.
Thus, there is a great opportunity and also a great challenge in picking different DNN models and system-resource settings to satisfy inference latency, accuracy,
and energy requirements.

\subsection{Understanding Variability}
\label{Sec:back_var} 
\begin{figure} [t!]
    \centering
    \includegraphics[width=0.88\linewidth]{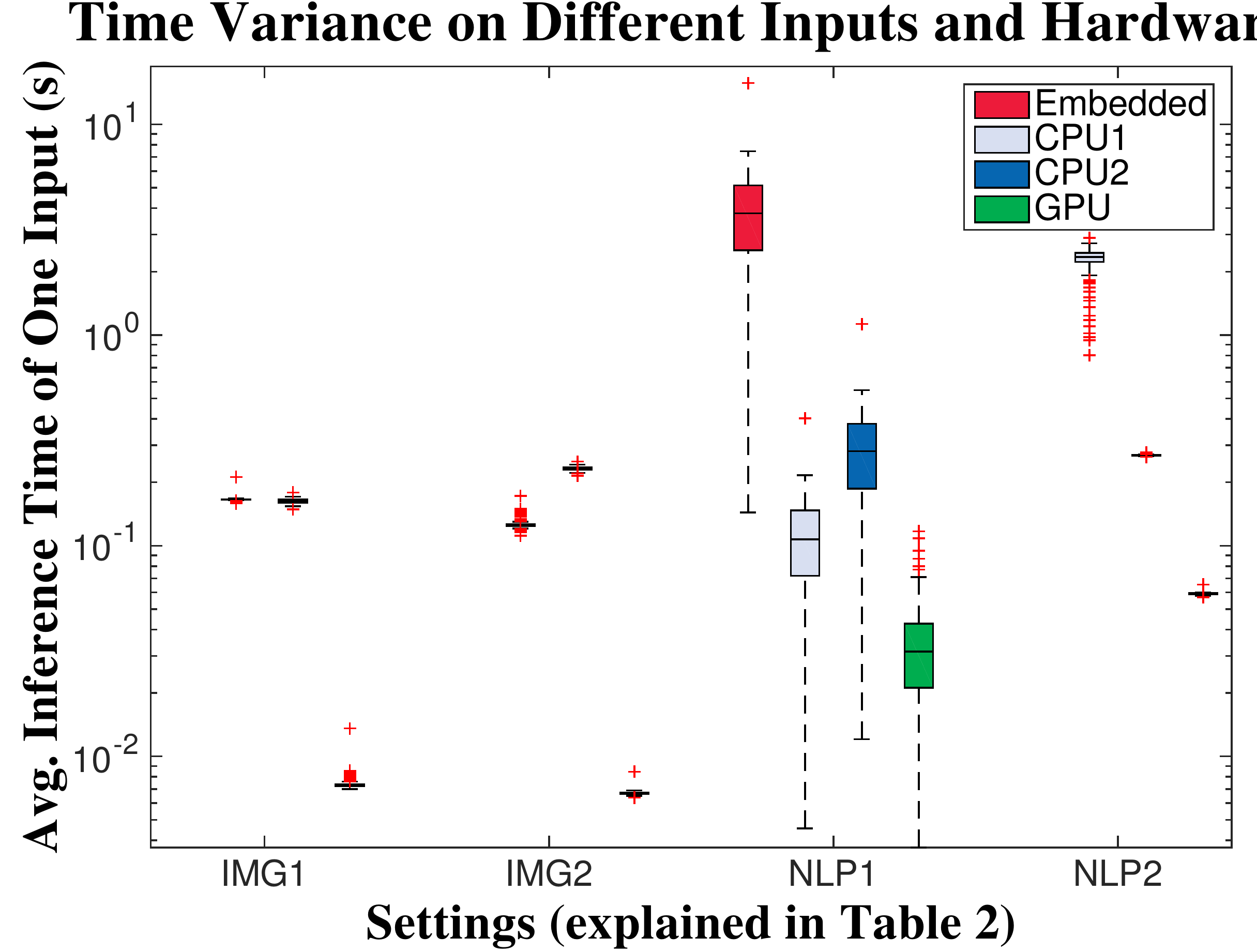}
    \vspace{-0.1in}
    \caption{Latency variance across inputs for different tasks and hardware
	   (Most tasks have 3 boxplots for 3 hardware platforms, CPU1-2, GPU from left to right; NLP1 has an extra boxplot for Embedded; other tasks run out of memory on Embedded; every box shows the 25th--75th percentile; points beyond the whiskers are >90th or <10th).
}
    \label{fig:variance}
\end{figure}

\begin{figure}[t!] 
    \centering
    \includegraphics[width=0.88\linewidth]{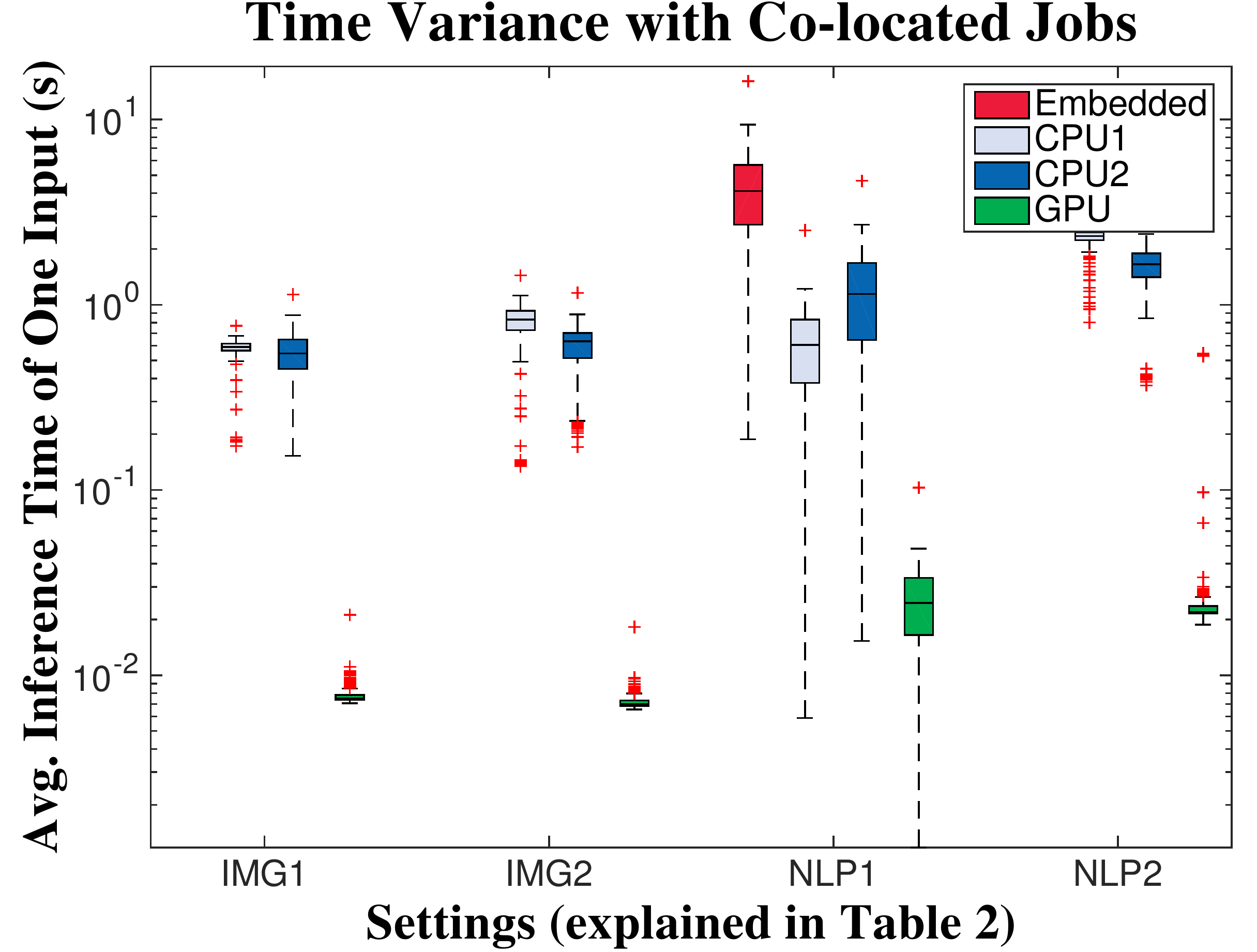}
    \vspace{-0.12in}
    \caption{Latency variance with co-located jobs
    (the memory-intensive STREAM benchmark \cite{stream} co-located on Embedded, CPU1-2; GPU-intensive Backprop \cite{rodinia} co-located on GPU)
    }
    \label{fig:Colocate_Memory_intensive2}
\end{figure}

To understand how DNN-inference
varies across inputs, platforms, and run-time environment and hence how (not) helpful is off-line
profiling, we run a set of experiments below, where
we feed the network one input at a time and use 1/10 of the total data for warm up, to emulate real-world scenarios. 
We plot the inference latency without and with
co-located jobs in Figure \ref{fig:variance} and
\ref{fig:Colocate_Memory_intensive2}, and we see several trends.

First, deadline violation is a realistic concern. 
Image classification on video has deadlines ranging from 1 second to the camera latency (e.g., 1/60 seconds) \cite{jiang2018chameleon}; the two NLP tasks, have
deadlines around 1 second
\cite{nielsen1994usability}. There is clearly no single 
inference task that meets all deadlines on all hardware.

Second, the inference variation among inputs is
relatively small particularly when there are no
co-located jobs (Fig. \ref{fig:variance}), 
except for that in NLP1, where
this large variance is mainly caused by different input lengths. For other tasks, outlier inputs exist but are rare.

Third, the latency and its variation across inputs
are both greatly affected by resource contention.
Comparing 
Figure \ref{fig:Colocate_Memory_intensive2} 
with Figure \ref{fig:variance}, we can see that the co-located job has increased both the median latency,
the tail inference, and the difference between these two for all tasks on all platforms. This trend also applies to other contention cases.

While the discussion above is about latency, similar conclusions apply to inference accuracy and energy: the accuracy typically drops to close to 0 when the inference time exceeds the latency requirement, and the energy consumption naturally changes with inference time.

\vspace{3pt}
\noindent {\bf Summary:} Deadline violations are realistic concerns and inference latency varies greatly across platforms, under  contention, and sometimes across inputs. Clearly,
sticking to one static DNN design across platforms and workloads leads to an unpleasant trade-off: always meeting the deadline by sacrificing accuracy or energy in most settings, or achieving a high accuracy some times but exceeding the deadline in others. Furthermore, it is also sub-optimal to make
run-time decisions based solely on off-line profiling, considering the variation caused by run-time contention.

\subsection{Understanding Potential Solutions}
\label{sec:back_pot}


We now show how confining adaptation to a single layer (just application or system) is insufficient.
We run the ImageNet classification on \emph{CPU1}.   
We examine a range of latency (0.1s-0.7s) and accuracy constraints (85\%-95\%), and try 
meeting those constraints while minimizing energy by either (1) configuring just the DNN (selecting a DNN from a family, like that in Figure \ref{fig:tradeoff_energy}) or (2) configuring just the system (by selecting resources to control energy--latency tradeoffs as in Figure \ref{fig:tradeoff_energy_latency}).  We compare these single-layer approaches to one that simultaneously picks the DNN and system configuration.  
As we are concerned with the ideal case, we create oracles by running 90 inputs in all possible DNN and system configurations, from which we find the best configuration for each input.  The App-level oracle uses the default system setting. The Sys-level oracle uses the default (highest accuracy) DNN.  

\begin{figure}
    \centering
    \includegraphics[width=0.88\linewidth]{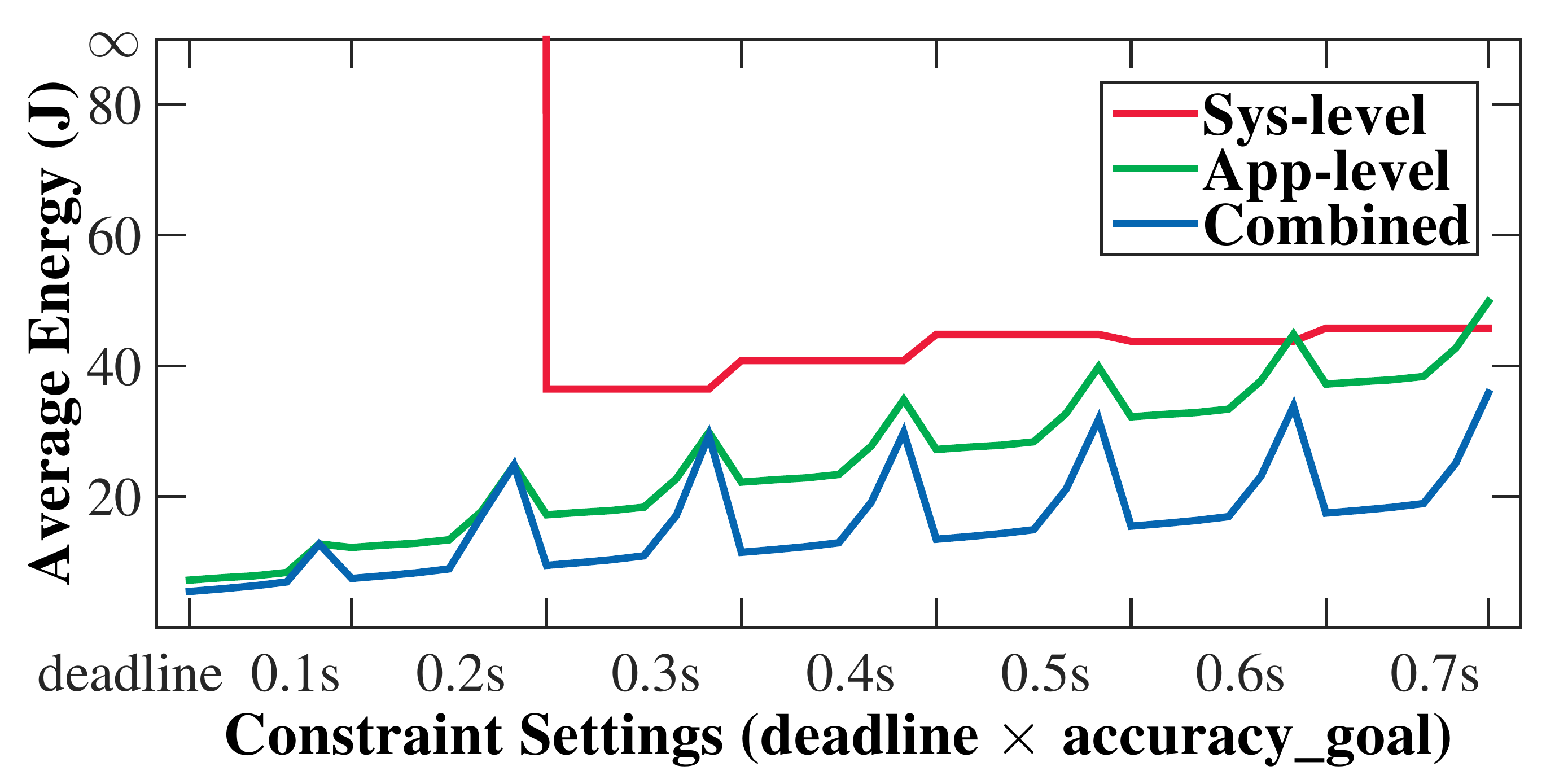}
    \vspace{-0.15in}
    \caption{Minimize energy task with latency and accuracy constraint @ CPU1. ($\infty$ means unable to meet the constraints)
    }
    \label{fig:motivation_oracle}
\end{figure}

Figure \ref{fig:motivation_oracle} shows the results.  As we have a three dimensional problem---meeting accuracy and latency constraints with minimal energy---we linearize the constraints and show them on the x-axis (accuracy is faster changing, with latency slower, so each latency bin contains all accuracy goals).  
There are several important conclusions here.  First, the App-only approach meets all possible accuracy and latency constraints, while the Sys-only approach cannot meet any constraints below 0.3s. Second, across the entire constraint range, App-only consumes significantly more energy than Combined (60\% more on average).
The intuition behind Combined's superiority is that there are discrete choices for DNNs; so when one is selected, there are almost always energy saving opportunities by tailoring resource usage to that DNN's needs.

\textbf{Summary:} Combining DNN and system level approaches achieves better outcomes. If left solely to the application, energy will be wasted.  If left solely to the system, many achievable constraints will not be met.  

\section{\Tool Run-time Inference Management}
\label{sec:scheduler}

\Tool's runtime system navigates the large tradeoff space created by \textit{combining} DNN-level and system-level adaptation.  \Tool meets 
user-specified latency, accuracy, and energy constraints and optimization goals while accounting for run-time
variations in environment or the goals themselves.

\subsection{Inputs \& Outputs of \Tool}
\Tool's inputs are specifications about (1) the adaption options, including a set of DNN models
$\mathbb{D} = \{d_i \mid i=1 \cdots K\}$ 
and a set of system-resource settings, expressed as  
different power-caps $\mathbb{P} = \{P_j \mid j=1 \cdots L\}$; 
and (2) the user-specified
requirements on latency, accuracy, and energy usage,
which can take the form of meeting constraints in any two of
these three dimensions while optimizing the third.
\Tool's output is the DNN model $d_i \in \mathbb{D}$
and the system-resource setting $p_j \in \mathbb{P}$ for the next inference-task input.

Formally, \Tool selects a DNN $d_i$ and a system-resource
setting $p_j$ to 
fulfill \textit{either}   
 of these user-specified goals.
 
     Maximizing inference accuracy $q$ (minimizing error) for an energy budget $\textbf{E}_\text{goal}$ and inference deadline $\textbf{T}_\text{goal}$:
    
\begin{equation}
\begin{aligned}
     \arg\max_{i,j} {q}_{i,j} \,\,\,\text{    s.t. } 
    e_{i,j} \leq \textbf{E}_{\text{goal}} \land t_{i,j} \leq \textbf{T}_{\text{goal}} 
 \end{aligned}
 \label{eq:max_acc}
\end{equation}

    Minimizing the energy use $e$ for an accuracy goal $\textbf{Q}_\text{goal}$ and inference deadline $\textbf{T}_\text{goal}$:
\begin{equation}
\begin{aligned}
      \arg\min_{i,j} e_{i,j} \,\,\,\text{    s.t. } 
     {q}_{i,j} \geq \textbf{Q}_{\text{goal}} \land t_{i,j} \leq \textbf{T}_{{goal}}
\end{aligned}
\label{eq:min_energy}
\end{equation}


We omit the discussion of meeting energy and accuracy constraints while minimizing latency as it is a trivial extension of the discussed techniques and we believe it to be the least practically useful.
We also omit the problem of optimizing all three dimensions, as it creates a feasibility problem, leaving nothing for optimization---lowest latency and highest accuracy are impractical to achieve simultaneously.

\paragraph{Generality}
Along the DNN-adaptation side, the input DNN set can consist
of any DNNs that offer different accuracy, latency, and
energy tradeoffs; e.g., those in Figure \ref{fig:tradeoff_energy_latency}.
In particular, \Tool can work with either or both of the broad classes of DNN adaptation approaches that have arisen recently, including: (1) traditional DNNs where the adaptation option should be selected prior to starting an inference task \cite{figurnov2017spatially, mcgill2017deciding, andreas2017convolutional, wu2018blockdrop,NestDNNMobicom} and (2) anytime DNNs that produce a series of outputs as they execute \cite{larsson2016fractalnet, branchynet, huang2017multi,lee2018anytime,wang2018anytime,hu2019learning,Our-icml-paper}. These two classes are similar in that they both vary things like the network depth or width to create latency/accuracy tradeoffs.  

On the system-resource side, \Tool uses a \emph{power cap} as the proxy to
system resource usage. Since both hardware \cite{rapl} and software resource managers \cite{PUPiL,PCP,packandcap} can convert power budgets into optimal performance resource allocations, \Tool is compatible with many different schemes from both commercial products and the research literature.

\subsection{\Tool Workflow}
\label{sec:steps}
 
\Tool works as a feedback controller.
It follows four steps to pick the DNN and resource
settings for each input $n$:

1) {Measurement.} \Tool records the processing time, energy usage, and computes inference accuracy for $n-1$.
    
2) {Goal adjustment.} 
\Tool updates the time goal $T_\text{goal}$ if necessary, considering the potential latency-requirement variation
across inputs. 
In some inference tasks,
a set of inputs share one
combined requirement
(e.g., in the NLP1 task 
in Table \ref{tbl:task}, all the words in 
a sentence are processed by a DNN one by one and
share one sentence-wise deadline)
and hence delays in previous input processing could greatly shorten the available time for the next input \cite{apollo,autoware}. Additionally, \Tool sets the goal latency to compensate for its own, worst-case overhead so that \Tool itself will not cause violations.

3)    {Feedback-based estimation.} \Tool 
computes the expected latency, accuracy, and energy consumption for every combination of DNN model and power setting.

4)    {Picking a configuration.} \Tool feeds all the updated estimations of latency, accuracy, and energy into Eqs. \ref{eq:max_acc} and \ref{eq:min_energy}, and gets the desired DNN model and power-cap setting for $n$.

The key task is step 3: the estimation needs to be
accurate and fast. In the remainder of this section,
we discuss key ideas and the
exact algorithm of our feedback-based estimation.






\subsection{Key Ideas of \Tool Estimation}
\label{sec:est_idea}

\textbf{Strawman} Solving Eqs. \ref{eq:max_acc} and \ref{eq:min_energy} would be trivially easy if the deployment environment is guaranteed to match the training and profiling environment: we could estimate $t_{i,j}$ to be the average (or worst case, etc) inference time 
$t_{i,j}^{\text{prof}}$  over a set of profiling inputs under model $d_i$ and power setting $p_j$.  
However, this approach does not work given 
the dynamic  input, contention, and requirement variation.

Next,
we present the key ideas behind how \Tool estimates
the inference latency, accuracy, and energy consumption under model $d_i$ and power setting $p_j$.


\textbf{How to estimate the inference latency $t_{i,j}$?}
To handle the run-time variation, 
a potential solution is to apply an estimator, like a Kalman filter \cite{kalman_filter}, to make dynamic predictions based on recent 
history about inferences under model $d_i$ and power $p_j$. 
The problem is that most models and power settings will not have
been picked recently and hence would have no recent history to feed into the estimator. This problem is a direct example of the challenge imposed by the large space of combined application and system options.

\textbf{Idea 1: Handle the large selection space with a single scalar value.}  
To make effective online estimation for {\it all} combinations of models and power settings,
 \Tool introduces a {\it global slow-down factor} $\xi$ to capture how the current 
environment differs from the profiled environment (e.g., due to co-running
processes, input variation, or other changes). Such an environmental slow-down
factor is independent from individual model or power selection. It can  fully leverage execution history, no matter which models
and power settings were recently used; it can then be used to estimate
$t_{i,j}$ based on $t_{i,j}^{\text{prof}}$ for all $d_i$ and $p_j$ combinations. 

Applying a \textit{global} slowdown factor for \textit{all} combinations of application and system-level settings is crucial for \Tool to make quick decisions for every inference task. Although it is possible that some perturbations may lead to different slowdowns for different configurations, the slight loss of accuracy here is out-weighed by the benefit of having a simple mechanism that allows prediction even for configurations that have not been used recently.

This idea is also novel for \Tool, as previous cross-stack management systems 
all use much more complicated models to estimate and select different setting combinations (e.g., using model predictive control to estimate combinations of settings \cite{mpcFSE2017}). 
\Tool's global slowdown factor is based on several unique features of DNN families that accomplish the same task with different accurarcy/latency tradeoffs.  We categorize these features as: (1) similarity of code paths and (2) proportionality of structure.  The first is based on the observation that DNNs do not have complex conditional code dependences, so we do not need to worry about the case where different inputs would exercise very different code paths.  Thus, what \Tool learns about latency, accuracy, and energy for one input will always inform it about future inputs.  The second feature refers to the fact that as DNNs in a family scale in latency, the proportion of different operations tend to be similar, so what \Tool learns about one DNN in the family generally applies to other DNNs in the same family. These properties of DNNs do not hold for many other types of software, where different inputs or additional functionality can invoke entirely different code paths, with different resource requirements or responses.


\textbf{How to estimate the accuracy under a deadline?} 
Given a deadline $\textbf{T}_{\text{goal}}$, the inference accuracy delivered by model $d_i$ and power setting $p_j$ is determined by three factors, as shown in Eq. \ref{eq:expected_acc}: (1) whether the inference result, 
which takes time $t_{i,j}$, can be generated before the deadline $\textbf{T}_{\text{goal}}$; (2) if yes, 
the accuracy is determined by the model $d_i$;\footnote{Since
it could be infeasible to calculate
the exact inference accuracy at run time, \Tool uses 
the average training accuracy of the selected DNN model $d_i$,
denoted as $q_{i}$, as the
inference accuracy, as long as the inference computation
finishes before the specified deadline.}
(3) if not, the accuracy drops to that offered
by a backup result $q_{\text{fail}}$. For traditional DNN models, without any output at the deadline, a
random guess will be used and $q_{\text{fail}}$ will be much worse than $q_i$.
For anytime DNN models that output multiple results as they are ready, 
the backup result is the latest output
\cite{larsson2016fractalnet, branchynet, huang2017multi,lee2018anytime,wang2018anytime,hu2019learning,Our-icml-paper},
which 
we discuss more
in Section \ref{sec:design_any}.

\begin{equation}
     {q}_{i,j}[\textbf{T}_{\text{goal}}] = 
    \left\{
    \begin{aligned}
        q_i \,\,\,\,\,\,, &\text{    if }  {t}_{i,j} \leq \textbf{T}_{\text{goal}} \\ 
        q_{\text{fail}} \,\,\,\,\,\,, &\text{    otherwise }
    \end{aligned}
    \right.
    \label{eq:expected_acc}
\end{equation}

A potential solution to estimate accuracy $q_{i,j}$ at the deadline
$\textbf{T}_{\text{goal}}$
is to simply feed the estimated $t_{i,j}$ into Eq.
\ref{eq:expected_acc}. 
However, this simple approach fails to account for two issues.  First, while DNNs are generally well-behaved, significant tail effects are possible (see Figure \ref{fig:variance}).  Second, Eq. \ref{eq:expected_acc} is not linear, and is best understood as a step function, where a failure to complete inference by the deadline results in a worthless inference output ($q_{fail}$).  Combined, these two issues mean that for tail inputs, inference will produce a worthless result; i.e., accuracy is not proportional to latency, but can easily fall to zero for tail inputs.  The tail will, of course, be increased if there is any unexpected resource contention.  Therefore, the simple approach of using the mean latency prediction fails to account for the non-linear affects of latency on accuracy.

\textbf{Idea 2: handle the runtime variation and account for tail behavior}
To handle the run-time variability mentioned in Section \ref{sec:intro},
\Tool treats the execution time 
$t_{i,j}$ and
the global slow-down factor $\xi$ 
as {\it random variables} drawn from a normal distribution.
\Tool uses a recently proposed extension to the Kalman filter to adaptively update the noise covariance \cite{update_kalman}. While this extension was originally proposed to produce better estimates of the mean, a novel approach in \Tool is using this  covariance estimate as a measure of system volatility. \Tool uses this Kalman filter extension to predict not just the mean accuracy, but also the likelihood of meeting the accuracy requirements in the current operating environment.  Section \ref{sec:eval_details} shows the advantages of our extensions.

\textbf{How to minimize energy or satisfy energy constraints?}
Minimizing energy or satisfying energy constraints
 is complicated, 
as the energy is related to, but cannot be easily
calculated by, the complexity of the selected model
$d_i$ and the power cap $p_j$.
As discussed in Section \ref{Sec:back_var}, the energy consumption
includes both that used during the inference under a given model
$d_i$ and that used  during the inference-idle
period, waiting for the next input. Consequently, it is not
straightforward to decide which power setting to use.


\textbf{Idea 3.} \Tool leverages insights from previous research, 
which shows that energy for latency-constrained systems can be efficiently expressed as a mathematical optimization problem \cite{Kim2015,Heiser1,Heiser2,caloree}. These frameworks optimize energy by scheduling available configurations in time.  Time is assigned to configurations so that the average performance hits the desired latency target and the overall energy (including idle energy) is minimal.  The key is that while the configuration space is large, the number of constraints is small (typically just two).  Thus, the number of configurations assigned a non-zero time is also small (equal to the number of constraints) \cite{Kim2015}.  Given this structure, the optimization problem can be solved using a binary search over available configurations, or even more efficiently with a hash table \cite{caloree}.

The only difficulty applying prior work to \Tool is that prior work assumed there was only a single job running at a time, while \Tool assumes that other applications might contend for resources.  Thus, \Tool cannot assume that there is a single system-idle state that will be used whenever the DNN is not executing.  
To address this challenge, \Tool continually estimates the system power when DNN inference is idle (but other non-inference tasks might be active), $p_{DNNidle}$, transforming Eq. \ref{eq:max_acc} is transformed into:   
\begin{equation}
\begin{aligned}
     \arg\max_{i,j} {q}_{i,j} [\textbf{T}_{\text{goal}}] \,\,\,\text{    s.t. } 
    p_{i,j}\cdotp t_{i,j} + p_{DNNidle} \cdotp t_{DNNidle} \leq \textbf{E}_{\text{goal}}
 \end{aligned}
 \label{eq:max}
\end{equation}



\subsection{\Tool Estimation Algorithm}
\label{sec:detailed_algorithm}


\textbf{Global Slow-down Factor $\xi$.} As discussed in Idea-1,
\Tool uses ${\xi}$
to reflect how the run-time environment differs from the profiling environment. Conceptually, if the inference task under model $d_i$ and power-cap $p_j$ took time $t_{i,j}$ at run time and took $t_{i,j}^{\text{prof}}$ on average to finish during profiling, the corresponding $\xi$ would be ${t_{i,j}}/{t_{i,j}^{prof}}$. \Tool estimates $\xi$ using recent execution history under any model or power setting.


Specifically, after an input $n-1$, \Tool computes $\xi^{(n-1)}$ as the ratio of the observed time  $t^{(n-1)}_{i,j}$ to the profiled time
$ t_{i,j}^{\text{prof}} $, and then uses a Kalman Filter\footnote{ A Kalman Filter is an optimal estimator that assumes a normal distribution and estimates a varying quantity based on multiple potentially noisy observations \cite{kalman_filter}.} to estimate the mean $\mu^{(n)}$ and variance $(\sigma^{(n)})^2$ of $\xi^{(n)}$ at input $n$.  
\Tool's formulation is defined in Eq. \ref{eq:kalman}, where $K^{(n)}$ is the Kalman gain variable; $R$ is a constant reflecting the measurement noise; $Q^{(n)}$ is the process noise capped with $Q^{(0)}$. We set a forgetting factor of process variance $\alpha=0.3$ \cite{update_kalman}. \Tool initially sets $K^{(0)}=0.5$, $R=0.001$, $Q^{(0)} = 0.1$, $\mu^{(0)}=1$, $(\sigma^{(0)})^2=0.1$, following the standard convention  \cite{kalman_filter}. \useshortskip
\begin{equation}
\vspace{-0.05in}
\left\{
\begin{aligned}
  Q^{(n)} &= \max \{ Q^{(0)}, \alpha Q^{(n-1)} + (1 \text{-} \alpha) ( K^{(n-1)} y^{(n-1)} )^2 \}\\
  K^{(n)} &= \frac{(1-K^{(n-1)})(\sigma^{(n-1)})^2 + Q^{(n)}}{(1-K^{(n-1)})(\sigma^{(n-1)})^2 + Q^{(n)} + R}  \\
 y^{(n)} &= {t^{(n-1)}_{i,j}}/{t^{\text{prof}}_{i,j}} - \mu^{(n-1)}\\
 \mu^{(n)} &= \mu^{(n-1)}+ K^{(n)} y^{(n)} \\
 (\sigma^{(n)})^2 &= (1-K^{(n-1)})(\sigma^{(n-1)})^2 + Q^{(n)}
\end{aligned}
\right.
\label{eq:kalman}
\end{equation}

Then, using $\xi^{(n)}$, \Tool estimates the inference time of input $n$ under any model $d_i$ and power cap $p_j$: $t_{i,j}^{(n)}$ = $\xi^{(n)} * t_{i,j}^{\text{prof}}$.


\textbf{Probability of meeting the deadline.}
Given the Kalman Filter estimation for the global slowdown factor, we can calculate  $Pr_{i,j}$ , the probability that the inference completes before the deadline $T_{goal}$. \Tool computes this value using a cumulative distribution function (CDF) based on the normal distribution of $\xi^{(n)}$ estimated by the Kalman Filter:
\begin{equation}
\begin{aligned}
Pr_{i,j}&=Pr[\xi^{(n)} \cdotp t_{i,j}^{\text{prof}}\leq T_{goal}] 
=  CDF(\xi^{(n)} \cdotp t_{i,j}^{\text{prof}}, T_{goal})\\
&=  CDF(\mu^{(n)}\cdotp t_{i,j}^{\text{prof}},\sigma^{(n)}, T_{goal})\\
\end{aligned}
\label{eq:probability}
\end{equation}

\textbf{Accuracy.}
As discussed in Idea-2, \Tool computes the estimated inference accuracy  $\hat{q}_{i,j}[\textbf{T}_{{\text{goal}}}]$  by considering $t_{i,j}$ as a random variable that follows normal distribution with its mean and variance computed based on that of $\xi$. Here $q_{i,j}$ represents the inference accuracy when the DNN inference finishes before the deadline, and $q_{fail}$ is the accuracy of a random guess:

\begin{equation}
\begin{aligned}
    \hat{q}_{i,j}[\textbf{T}_{{goal}}] 
    =& E({q}_{i,j}[\textbf{T}_{{goal}}] \mid t_{i,j}^{(n)}) \\
    =& E({q}_{i,j}[\textbf{T}_{{goal}}] \mid \xi^{(n)} \cdotp t_{i,j}^{\text{prof}}) \\
    =& Pr_{i,j} \cdotp {q}_{i,j} + (1-Pr_{i,j}) \cdotp {q}_{fail} \\
    \xi^{(n)} \sim& \mathcal{N} (\mu^{(n)},\,(\sigma^{(n)})^{2})
\end{aligned}
\label{eq:estacc}
\end{equation}

\textbf{Energy.} 
As discussed in Idea-3, \Tool predicts energy consumption by separately estimating energy during (1) DNN execution: estimated by multiplying the power limit by the estimated latency and (2) between inference inputs: estimated based on the recent history of inference idle power using the Kalman Filter in Eq. \ref{eq:kalman_energy}.
$\phi^{(n)}$ is the predicted DNN-idle power ratio, $M^{(n)}$ is process variance, $S$ is process noise, $V$ is measurement noise, and $W^{(n)}$ is the Kalman Filter gain.  \Tool initially sets $M^{(0)}=0.01$, $S=0.0001$, $V=0.001$.

\begin{equation}
\left\{
\begin{aligned}
   W^{(n)} &= \frac{M^{(n-1)} + S}{M^{(n-1)} + S + V}  \\
   M^{(n)} &= (1-W^{(n)})(M^{(n-1)} + S) \\
   \phi^{(n)} &= \phi^{(n-1)}+ W^{(n)}({p_\text{idle}}/{p^{(n-1)}_{i,j}} - \phi^{(n-1)})
\end{aligned}
\right.
\label{eq:kalman_energy}
\end{equation}

\Tool then predicts the energy by Eq. \ref{eq:energy_prediction}. Unlike equation \ref{eq:estacc} that uses probabilistic estimates, energy estimation is calculated without the notion of probability. The inference power is the same whenever the inference misses and meets the deadline because \Tool sets power limits. Therefore it is safe to estimate the energy by its mean without considering the distribution of its possible latency. See Eq. \ref{eq:energy_user} to estimate energy by its worst case latency percentile. 

\begin{equation}
\begin{aligned}
    e^{(n)}_{i,j} = p_{i,j} \cdotp \xi^{(n)} \cdotp  t^{\text{prof}}_{i,j} + \phi^{(n)} \cdotp p_{i,j} \cdotp (\textbf{T}_{goal}-(\xi^{(n)} \cdotp  t^{\text{prof}}_{i,j}))
\end{aligned}
\label{eq:energy_prediction}
\end{equation}

\textbf{Selecting Configurations.}
Given the estimates of latency, expected accuracy, and energy consumption, \Tool generates a set of valid configurations which meets all of the constraints. \Tool then chooses the best valid configuration according to the optimization goal; i.e., \Tool selects a configuration that solves either Eq. \ref{eq:max_acc} or   Eq. \ref{eq:min_energy} using the estimated latency, accuracy, and energy from  Equations \ref{eq:kalman}, \ref{eq:estacc}, and \ref{eq:energy_prediction}, respectively.

In unpredictable environments---i.e., those with high estimated variance from Eq. \ref{eq:kalman}---\Tool is to be more conservative, selecting from fewer valid configurations. Consider an example scenario with two DNN candidates. The larger one has an estimated accuracy of 0.98, and 97\% probability to meet the deadline. Meanwhile, the smaller one has 0.95 estimated accuracy and 99.9\% probability, respectively. The larger DNN has lower probability because it takes longer time. When observed variance is low, \Tool picks the larger DNN for its higher expected accuracy (i.e., $97\% \times 0.98 = 0.951$ compared with the smaller one's $99.9\% \times 0.95 = 0.949$). When variance is high, however, the Kalman Filter has difficulty in predicting the latency, and the estimate will be different from the measured value. This increases the Kalman Filter's $Q$ value (Eq. \ref{eq:kalman}, and thus increases its estimated variance ($\sigma^2$). The higher estimated variance means that the probability of completion by the deadline for all configurations in equation \ref{eq:probability} will be decreased. Consequently, the probability of selecting a larger DNN will be decreased more than that of the small DNN because it has larger latency. In our example, the larger DNN's probability of completion drops to 95\% from 97\%, thus decreasing the expected accuracy to 0.941. In contrast the smaller DNN only drops its probability to 99.5\% from 99.9\%, decreasing its expected accuracy to 0.945. \Tool then chooses smaller DNN which now has a higher expected accuracy (because it is more likely to complete) under high variance environment.

\textbf{Manipulating \Tool's Probabilistic Guarantees.}
\Tool default setting is using a full mathematical expectation without explicitly defining a probabilistic threshold ($Pr_{th}$), which represents the probability of meeting the constraints. Users can set this probabilistic threshold ($Pr_{th}$) according their needs and then \Tool will not select configurations for which the probability is below this threshold. Adding this capability is as simple as adding another constraint to Eq. \ref{eq:max_acc} in the maximizing accuracy scenario:
    
\begin{equation}
\begin{aligned}
     \arg\max_{i,j} {q}_{i,j} \,\,\,\text{    s.t. } 
    e_{i,j} \leq \textbf{E}_{\text{goal}} \land t_{i,j} \leq \textbf{T}_{\text{goal}} \land Pr_{i,j} \geq Pr_{th} 
 \end{aligned}
 \label{eq:guarantee_max_acc}
\end{equation}

In minimizing energy scenario, Eq. \ref{eq:min_energy} is modified to be:
\begin{equation}
\begin{aligned}
      \arg\min_{i,j} e_{i,j} \,\,\,\text{    s.t. } 
     {q}_{i,j} \geq \textbf{Q}_{\text{goal}} \land t_{i,j} \leq \textbf{T}_{{goal}} \land Pr_{i,j} \geq Pr_{th}
\end{aligned}
\label{eq:guarantee_min_energy}
\end{equation}

Energy estimation can also be updated accordingly for users who want more control over \Tool's energy guarantees:

\begin{equation}
\begin{aligned}
    e^{(n)}_{i,j} =& p_{i,j} \cdotp CDF^{-1}(\xi^{(n)} \cdotp  t^{\text{prof}}_{i,j},Pr_{th})+\\
    & \phi^{(n)} \cdotp p_{i,j} \cdotp (\textbf{T}_{goal}-CDF^{-1}(\xi^{(n)} \cdotp  t^{\text{prof}}_{i,j},Pr_{th})),
\end{aligned}
\label{eq:energy_user}
\end{equation}
where $CDF^{-1}(\xi^{(n)} \cdotp  t^{\text{prof}}_{i,j},Pr_{th})$ is the inverse of cumulative distribution function. It takes two inputs: (1) the distribution function of random variable $\xi^{(n)} \cdotp  t^{\text{prof}}_{i,j}$ and (2) the user threshold $Pr_{th}$ which indicates the probability of meeting the goal. It outputs the predicted latency such that it is the worst case latency of $Pr_{th}$ percentile from distribution $t_{i,j}$. Compared with Eq. \ref{eq:energy_prediction}, the energy estimation by this equation will be higher as it uses a higher percentile latency. Thus, \Tool will reject more configurations and may lead to lower expected accuracy as the cost of tighter energy bounds.

\subsection{Integrating \Tool with Anytime DNNs}
\label{sec:design_any}

An anytime DNN is an inference model that outputs
a series of increasingly accurate inference results---$o_1$,
$o_2$, ... $o_k$, with $o_t$ more reliable than $o_{t-1}$.
A variety of recent works \cite{larsson2016fractalnet, branchynet, huang2017multi,lee2018anytime,wang2018anytime, Our-icml-paper} have proposed DNNs supporting anytime inference, covering a variety of problem domains. 
\Tool easily works with not only traditional DNNs
but also Anytime DNNs. 
The only change is that
$q_{\text{fail}}$ in Eq. \ref{eq:expected_acc} no longer corresponds
to a random guess. That is, when the inference could not generate
its final result $o_k$ by the deadline
$\textbf{T}_{\text{goal}}$, an earlier result $o_x$ can be
used with a much better accuracy than that of a random guess.
The updated accuracy equation is below:
\begin{equation}
    q_{.,j} = 
    \left\{
    \begin{aligned}
        q_k \,\,\,\,\,\,, &\text{    if }  {t}_{k,j} \leq \textbf{t}_{\text{goal}} \\ 
        q_{k-1} \,\,\,\,\,\,, &\text{    if } t_{k-1,j} \leq \textbf{t}_{\text{goal}}<{t}_{k,j} \\
        & \cdots 
        \\
        q_{\text{fail}} \,\,\,\,\,\,, &\text{    otherwise }
    \end{aligned}
    \right.
    \label{eq:expected_acc_any}
\end{equation}

Existing anytime DNNs consider latency but not energy constraints---an anytime DNN will keep running until the latency deadline arrives and the last output will be delivered to the user. \Tool naturally improves Anytime DNN energy efficiency, stopping the inference sometimes before the deadline based on its estimation to meet not only latency and accuracy, but also energy requirements. 

Furthermore, \Tool can work with a set of traditional DNNs and an Anytime DNN together to achieve the best combined result.
The reason is that Anytime DNNs generally sacrifice accuracy for flexibility. When we feed a group of traditional DNNs and one Anytime DNN to construct the  candidacy set $\mathbb{D}$, with Eq. \ref{eq:estacc}, \Tool naturally selects the Anytime DNN when the environment is changing rapidly (because the expected accuracy of an anytime DNN will be higher given that variance), and the regular DNN, which has slightly higher accuracy with similar computation, when it is stable, getting the best of both worlds. 

In our evaluation, we will use the nested design from \cite{Our-icml-paper}, which provides a generic coverage of anytime DNNs. 

\subsection{Limitations and Discussions} 
\label{sec:design_limit}

\textbf{Assumptions of the Kalman Filter.}
\Tool's prediction, particularly the Kalman Filter, relies on the feedback from recent input processing. Consequently, it requires at least one input to react to sudden changes. 
Additionally, the Kalman filter formulations assume that the underlying distributions are normal, which may not hold in practice.  If the behavior is not Gaussian, the Kalman filter will produce bad estimations for the mean of $\xi$ for some amount of time.

Having said that, as will be shown by our experiments, no single distribution fits all 
real-world scenarios and normal distribution is the best fit 
we can find in practice (Figure \ref{fig:distribution}). Furthermore, \Tool is specifically designed to handle deviation from the normal-distribution assumption, novelly using the Kalman Filter’s covariance estimation to measure system volatility and accounting for volatility in the accuracy/energy estimations. 
Consequently, after just 2--3 such bad predictions of means, the estimated variance will increase, which will then trigger \Tool to pick anytime DNN over traditional DNNs or pick a low-latency traditional DNN over high-latency ones, because the former has a better chance to produce results at latency
deadlines and hence a higher expected accuracy under high variance. So---worst case---\Tool will choose a DNN with slightly less accuracy than what 
could have been used with the right model of randomness. Users can also compensate for extremely aberrant latency distributions by increasing the value of $Q^{(0)}$ in Eq. \ref{eq:kalman}. 
As we will see in Section \ref{sec:eval_details}, \Tool performs well even when the 
distribution is not normal.

\textbf{Probabilistic guarantees.}
\Tool provides probabilistic, not hard, guarantees.  As \Tool estimates not just average timing, but the distributions of possible timings, it can provide arbitrarily many nines of assurance that it will meet latency or accuracy goals but cannot provide 100\% guarantee. Providing 100\% guarantees requires the information of worst case execution time (WCET), a latency value that guarantees there is no slower latency with probability of 1. \Tool does not assume the availability of such information and hence cannot provide hard guarantees \cite{realtime-book}. 

\textbf{Safety guarantees.} While \Tool does not explicitly model safety requirements, it can be configured to prioritize accuracy over other dimensions.
In scenarios where users particularly value safety (e.g., auto-driving), they could set a high accuracy requirement or 
even remove the energy constraints. 

\textbf{Concurrent inference jobs.} \Tool is currently designed to support one inference job at a time. 
To support multiple concurrent inference jobs,
future work needs to extend \Tool to coordinate across
these concurrent jobs. We expect the main idea of \Tool,
such as using a global slowdown factor to estimate system
variation, to still apply.

\textbf{Scope of \Tool.}
Finally, how the inference behaves ultimately depends not
only on \Tool, but also on the DNN models and
system-resource setting options. As we will
evaluate in Section \ref{sec:experiment}, \Tool helps make the best use of supplied DNN models, 
but does not eliminate the difference between different 
DNN models.

\section{Implementation}
\label{sec:select_power}

We implement \Tool for both CPUs and GPUs.
On CPUs, \Tool adjusts power through Intel's RAPL interface \cite{rapl}, which allows software to set a hardware power limit. On GPUs, \Tool uses PyNVML to control frequency and builds a power-frequency lookup table. \Tool can also be applied to other approaches that translate power limits into settings for combinations of resources \cite{PCP,Bard,packandcap,PUPiL}.

In our experiments, \Tool considers a series of power settings within the feasible range with 2.5W interval on our test laptop and a 5W interval on our test CPU server and GPU platform, as the latter has a wider power range than the former. The number of power buckets is configurable.

\Tool incurs small overhead in both scheduler computation and switching from one DNN/power-setting to another, just 0.6--1.7\% of an input inference time. We explicitly account for overhead by subtracting it from the user-specified goal (see step 2 in Section \ref{sec:steps}).

Users may set goals that are not achievable.  If \Tool cannot meet all constraints, it prioritizes latency highest, then accuracy, then power. This hierarchy is configurable.

\section{Experimental Evaluation}
\label{sec:experiment}

\begin{table}
\footnotesize 
\setlength{\tabcolsep}{0.9mm}
\centering
\begin{tabular}{|l|l|l|l|l@{\hspace{0.1em}}|}
\hline
\rowcolor{lightgray}
     &\multicolumn{3}{l|}{Run-time environment setting}\\
\hline
Default & \multicolumn{3}{l|}{Inference task has no co-running process}\\
\multirow{2}{*}{Memory}  & \multicolumn{3}{l|}{Co-locate with memory-hungry STREAM \cite{stream} (@CPU)}\\ 
  & \multicolumn{3}{l|}{Co-locate with Backprop from Rodinia-3.1 \cite{rodinia} (@GPU)}\\
\multirow{2}{*}{Compute}     & \multicolumn{3}{l|}{Co-locate with Bodytrack from PARSEC-3.0 \cite{parsec} (@CPU)}\\
  & \multicolumn{3}{l|}{Co-locate with the forward pass of Backprop \cite{rodinia} (@GPU)}\\

\hline 
\rowcolor{lightgray}
                   & \multicolumn{3}{l|}{Ranges of constraint setting} \\
\hline
Latency         & \multicolumn{3}{l|}{0.4x--2x mean latency* of the largest Anytime DNN}\\
Accuracy        & \multicolumn{3}{l|}{Whole range achievable by trad. and Anytime DNN}\\  
Energy          & \multicolumn{3}{l|}{Whole feasible power-cap ranges on the machine}\\
\hline
\rowcolor{lightgray}
 Task & Trad. DNN & Anytime \cite{Our-icml-paper} & Fixed deadline?\\ 
\hline
Image Classifi.    &  Sparse ResNet & Depth-Nest & Yes \\
Sentence Pred.        & RNN             & Width-Nest & No \\ 
\hline
\rowcolor{lightgray}
Scheme ID & \multicolumn{2}{l|}{DNN selection} & Power selection\\ 
\hline
Oracle                &  \multicolumn{2}{l|}{Dynamic optimal} & Dynamic optimal \\
Oracle$_{\text{Static}}$&\multicolumn{2}{l|}{Static optimal}  & Static optimal  \\
\hline 
App-only  & \multicolumn{2}{l|}{One Anytime DNN}          &System Default  \\
Sys-only & \multicolumn{2}{l|}{Fastest traditional DNN}& State-of-Art\cite{poet} \\
No-coord & \multicolumn{2}{l|}{Anytime DNN w/o coord. with Power}& State-of-Art\cite{poet} \\
\hline 
ALERT                 & \multicolumn{2}{l|}{\Tool default}          & \Tool default \\
\ToolAny & \multicolumn{2}{l|}{\Tool w/o traditional DNNs} & \Tool default\\
\ToolTrad & \multicolumn{2}{l|}{\Tool w/o Anytime DNNs} & \Tool default\\
\hline 
\end{tabular}
\caption{Settings and schemes under evaluation \textmd{(* measured under
default setting without resource contention)}}
\label{tbl:schemes}
\end{table}


\def\colorModel{rgb}
\newcommand\ColCell[1]{
  \def \black{0.9}
  \def\xxx{1.19-#1/1.1}
   \IfSubStr{#1}{A} {\renewcommand{\xxx}{1}}{}
   \IfSubStr{#1}{O} {\renewcommand{\xxx}{1}}{}
   \IfSubStr{#1}{o} {\renewcommand{\xxx}{1}}{}
   \IfSubStr{#1}{0.} {\renewcommand{\xxx}{1}}{}
   \IfSubStr{#1}{1.} {\renewcommand{\xxx}{\black}}{}
   \IfSubStr{#1}{2.} {\renewcommand{\xxx}{\black}}{}
   \IfSubStr{#1}{3.} {\renewcommand{\xxx}{\black}}{}
   \IfSubStr{#1}{4.} {\renewcommand{\xxx}{\black}}{}
   \IfSubStr{#1}{5.} {\renewcommand{\xxx}{\black}}{}
   \IfSubStr{#1}{6.} {\renewcommand{\xxx}{\black}}{}
   \IfSubStr{#1}{7.} {\renewcommand{\xxx}{\black}}{}
   \IfSubStr{#1}{8.} {\renewcommand{\xxx}{\black}}{}
   \IfSubStr{#1}{9.} {\renewcommand{\xxx}{\black}}{}
   \IfSubStr{#1}{^} {\renewcommand{\xxx}{0.8}}{}

  \pgfmathparse{\xxx<0.56?0:1}  
    \ifnum\pgfmathresult=0\relax\color{white}\fi
  
  \pgfmathsetmacro\compA{\xxx}      
  \pgfmathsetmacro\compB{\xxx} 
  \pgfmathsetmacro\compC{\xxx}      
  \IfSubStr{#1}{A} {\relax\color{black}}{}
  \IfSubStr{#1}{O} {\relax\color{black}}{}
  \IfSubStr{#1}{o} {\relax\color{black}}{}
  \edef\x{\noexpand\centering\noexpand\cellcolor[\colorModel]{\compA,\compB,\compC}}\x #1
  } 
\newcolumntype{E}{>{\collectcell\ColCell}m{0.99cm}<{\endcollectcell}}  

\begin{table*}
\centering
\footnotesize{
\setlength{\tabcolsep}{1mm}
\begin{tabular}{|c|c|c||EEEEEE||EEEEEE|} 
\hline
Plat.                 & DNN                                                                       & Work. & ALERT & ALERT-Any & Sys-only     & App-only & No-coord     & Oracle & ALERT & ALERT-Any & Sys-only & App-only    & No-coord & Oracle  \\ 
\hline
\multicolumn{3}{|c||}{}                                                                                    & \multicolumn{6}{c||}{Energy in Minimizing Energy Task}                 & \multicolumn{6}{c|}{Error Rate in Minimizing Error Task}        \\ 
\hline
\multirow{6}{*}{CPU1} & \multirow{3}{*}{\begin{tabular}[c]{@{}c@{}}Sparse\\ Resnet \end{tabular}} & Idle  & 0.64  & 0.68      & $1.08^{19}$  & 1.19     & $0.94^{1}$   & 0.64   & 0.91  & 0.92      & 1.35     & $1.02^{3}$  & $0.91^{3}$  & 0.89    \\ 
\cline{3-15}
                      &                                                                           & Comp.   & 0.57  & 0.58      & $0.80^{19}$  & 1.30     & $1.39^{1}$   & 0.57   & 0.38  & 0.39      & 0.51     & $1.35^{24}$ & $0.39^{6}$  & 0.36    \\ 
\cline{3-15}
                      &                                                                           & Mem.  & 0.53  & 0.55      & $0.76^{19}$  & 1.43     & $1.37^{2}$   & 0.53   & 0.34  & 0.34      & 0.46     & $1.47^{28}$ & $0.39^{2}$  & 0.33    \\ 
\cline{2-15}
                      & \multirow{3}{*}{RNN}                                                      & Idle  & 0.61  & 0.65      & $1.01^{30}$  & 1.34     & $0.95^{2}$   & 0.61   & 0.87  & 0.87      & 0.87     & $0.87^{21}$ & $0.87^{14}$ & 0.86    \\ 
\cline{3-15}
                      &                                                                           & Comp.   & 0.60  & 0.57      & $0.93^{30}$  & 1.21     & $1.26^{5}$   & 0.60   & 0.42  & 0.44      & 0.50     & $0.46^{28}$ & $0.46^{23}$ & 0.42    \\ 
\cline{3-15}
                      &                                                                           & Mem.  & 0.54  & 0.56      & $0.95^{31}$  & 1.45     & $1.24^{9}$   & 0.54   & 0.45  & 0.45      & 0.50     & $0.57^{28}$ & $0.54^{27}$ & 0.44    \\ 
\hline
\multirow{6}{*}{CPU2} & \multirow{3}{*}{\begin{tabular}[c]{@{}c@{}}Sparse\\ Resnet \end{tabular}} & Idle  & 0.93  & 0.88      & $0.96^{20}$  & 0.99     & 1.18         & 0.91   & 0.68  & 0.68      & 0.97     & $0.79^{2}$  & $0.71^{24}$ & 0.66    \\ 
\cline{3-15}
                      &                                                                           & Comp.   & 0.59  & 0.57      & $0.60^{23}$  & 1.00     & 1.01         & 0.58   & 0.58  & 0.57      & 0.85     & $0.74^{16}$ & $0.71^{29}$ & 0.55    \\ 
\cline{3-15}
                      &                                                                           & Mem.  & 0.38  & 0.37      & $0.39^{19}$  & 0.65     & $0.63^{13}$  & 0.38   & 0.24  & 0.82      & 0.32     & $0.33^{17}$ & $0.75^{31}$ & 0.21    \\ 
\cline{2-15}
                      & \multirow{3}{*}{RNN}                                                      & Idle  & 0.87  & 0.99      & $0.80^{34}$  & 1.04     & $1.00^{6}$   & 0.83   & 0.84  & 0.85      & 0.99     & $0.89^{14}$ & $0.89^{1}$  & 0.84    \\ 
\cline{3-15}
                      &                                                                           & Comp.   & 0.60  & 0.60      & $0.55^{34}$  & 0.99     & $0.86^{7}$   & 0.60   & 0.51  & 0.52      & 0.60     & $0.53^{21}$ & $0.54^{17}$ & 0.52    \\ 
\cline{3-15}
                      &                                                                           & Mem.  & 0.52  & 0.51      & $0.43^{33}$  & 0.70     & $0.85^{14}$  & 0.52   & 0.26  & 0.27      & 0.31     & $0.28^{21}$ & $0.27^{17}$ & 0.26    \\ 
\hline
\multirow{3}{*}{GPU}  & \multirow{3}{*}{\begin{tabular}[c]{@{}c@{}}Sparse\\ Resnet \end{tabular}} & Idle  & 0.97  & 0.99      & $0.92^{20}$  & 1.36     & 1.37         & 0.92   & 0.90  & 0.92      & 1.22     & $1.09^{2}$  & $1.74^{12}$ & 0.86    \\ 
\cline{3-15}
                      &                                                                           & Comp. & 0.96  & 0.97      & $0.94^{20}$  & 1.66     & 1.77         & 0.89   & 0.32  & 0.34      & 1.28     & $1.21^{23}$ & $2.50^{18}$ & 0.30    \\ 
\cline{3-15}
                      &                                                                           & Mem. & 0.97  & 1.01      & $0.91^{20}$  & 1.39     & 1.43         & 0.91   & 0.89  & 0.92      & 1.22     & $1.11^{2}$  & $1.81^{14}$ & 0.86    \\ 

\hline
\rowcolor[rgb]{0.749,0.749,0.749} \multicolumn{3}{|c||}{Harmonic mean}                                     & 0.64  & 0.64      & $0.73^{27}$  & 1.11     & $1.08^{4}$   & 0.62   & 0.46  & 0.47      & 0.63     & $0.67^{16}$ & $0.63^{15}$ & 0.45    \\
\hline
\end{tabular}
\vspace{-0.1in}
\caption{Average energy consumption and error rate normalized to \emph{Oracle}$_\text{Static}$, smaller is better. (Each cell is averaged over 35--40  constraint settings; superscript means \# of constraint settings violated for more than 10\% of the time; those
settings' results are not part of the energy average.)}
\label{tbl:result_summary}
}
\end{table*}

We apply \Tool to different inference tasks
on both CPU and GPU with and without resource contention from co-located jobs.  We set \Tool to  (1) reduce energy while
satisfying latency and accuracy requirements and 
(2) reduce error rates while satisfying latency and
energy requirements. We  compare
\Tool with both oracle and state-of-the-art
schemes and evaluate detailed design decisions.


\subsection{Methodology}
\textbf{Experimental setup.}
We use the three platforms listed in
Table \ref{tbl:config}: \emph{CPU1}, \emph{CPU2}, and \emph{GPU}.
On each, we run inference tasks\footnote{For GPU, we only
run image classification task there, as the RNN-based
sentence prediction task is better suited for CPU \cite{zhang2018deepcpu}.},
image classification and sentence prediction, under three 
different resource-contention scenarios:

\begin{itemize}
\item No contention: the inference task is the only job running, referred to as ``Default'';

\item Memory dynamic: the inference task is running together with a memory-intensive job that repeatedly gets stopped and then started, representing dynamic memory resource contention, referred to as ``Memory''; 

\item Computation dynamic: the inference task is running together with a computation-intensive job that repeatedly gets stopped and then started, representing dynamic computation resource contention, referred to as ``Compute''. 
\end{itemize}

We then evaluate a number of management schemes' ability to meet latency, accuracy, and energy constraints. Table \ref{tbl:schemes} lists the details.

\textbf{Schemes under evaluation.}
We give \Tool three different DNN sets: traditional DNN models (\ToolTrad), an Anytime DNN (\ToolAny), and both (\Tool).

We compare with
two {\it Oracle}$_*$ schemes that have perfect predictions for every input under every DNN/power setting (i.e., impractical). The
``Oracle" allows DNN/power settings to change across inputs,
representing the best possible results; the ``Oracle$_\text{Static}$'' has one fixed setting across inputs, representing the best results without dynamic adaptation.

Finally, we compare with
three state-of-the-art approaches:

\begin{itemize}
\item ``App-only'' conducts adaptation only at the application level through an Anytime DNN \cite{Our-icml-paper};
\item  ``Sys-only'' conducts adaptation
only at the system level following an existing resource-management
system that minimizes energy under soft real-time constraints \cite{caloree}\footnote{Specifically, this adaptation uses a feedback scheduler that 
predicts inference latency based on Kalman
Filter.} and uses the fastest
candidate DNN to avoid latency violations; 
\item ``No-coord'' uses both the Anytime DNN for application-level adaptation \textit{and} the power-management scheme \cite{caloree} to adapt power, but with these two working independently. 
\end{itemize}

\subsection{Overall Results}

\begin{figure}
  \centering
  \includegraphics[width=.99\linewidth]{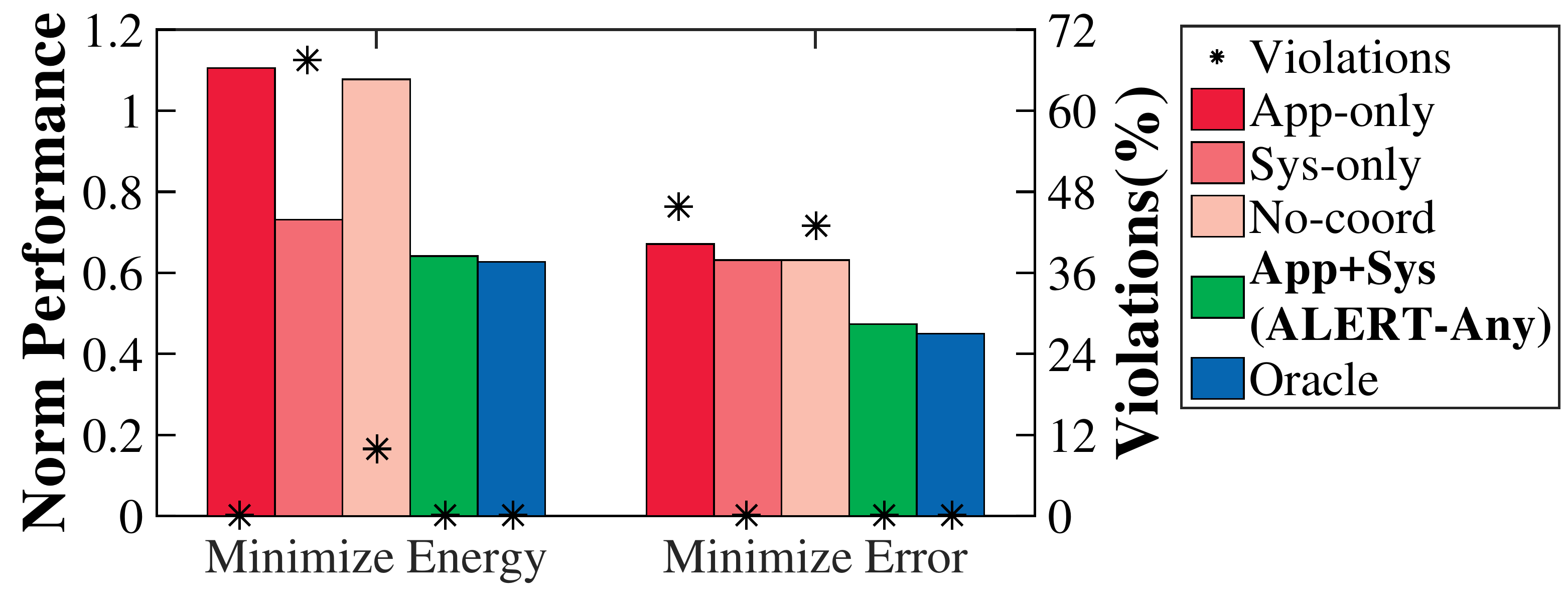}
  \vspace{-0.1in}
  \caption{Result Summary: average performance normalized to Oracle$_{\text{Static}}$. Violations\% refers to \%-of-constraint-settings under which a scheme incurs more than 10\% violation of all inputs. \textmd{(Smaller is better; Details in Table \ref{tbl:result_summary})}}
  \label{fig:scheduler_summary}
\end{figure}

Table \ref{tbl:result_summary} shows the results for all schemes for different tasks on different platforms and environments.
Each cell shows the average energy or accuracy under 35--40 combinations of latency, accuracy, and
energy constraints (the settings are detailed
in Table \ref{tbl:schemes}), normalized to the  {Oracle}$_\text{Static}$ result. Figure \ref{fig:scheduler_summary} compares these results, where lower bars represent
better results and lower *s represent fewer
constraint violations.
\Tool and 
\Tool$_\text{Any}$ both work very well for all settings.
They outperform 
state-of-the-art approaches, which
have a significant number of constraint violations, as visualized by
the many superscripts in Table \ref{tbl:result_summary} and
the high * positions in Figure \ref{fig:scheduler_summary}. \Tool outperforms
Oracle$_\text{Static}$ because it adapts to dynamic variations. \Tool  
also comes very close to the
theoretically optimal Oracle.

\begin{figure*}[t] 
  \centering
  \vspace{-0.1in}
  \subfloat[CPU1, Image Classification]{\includegraphics[width=.245\linewidth]{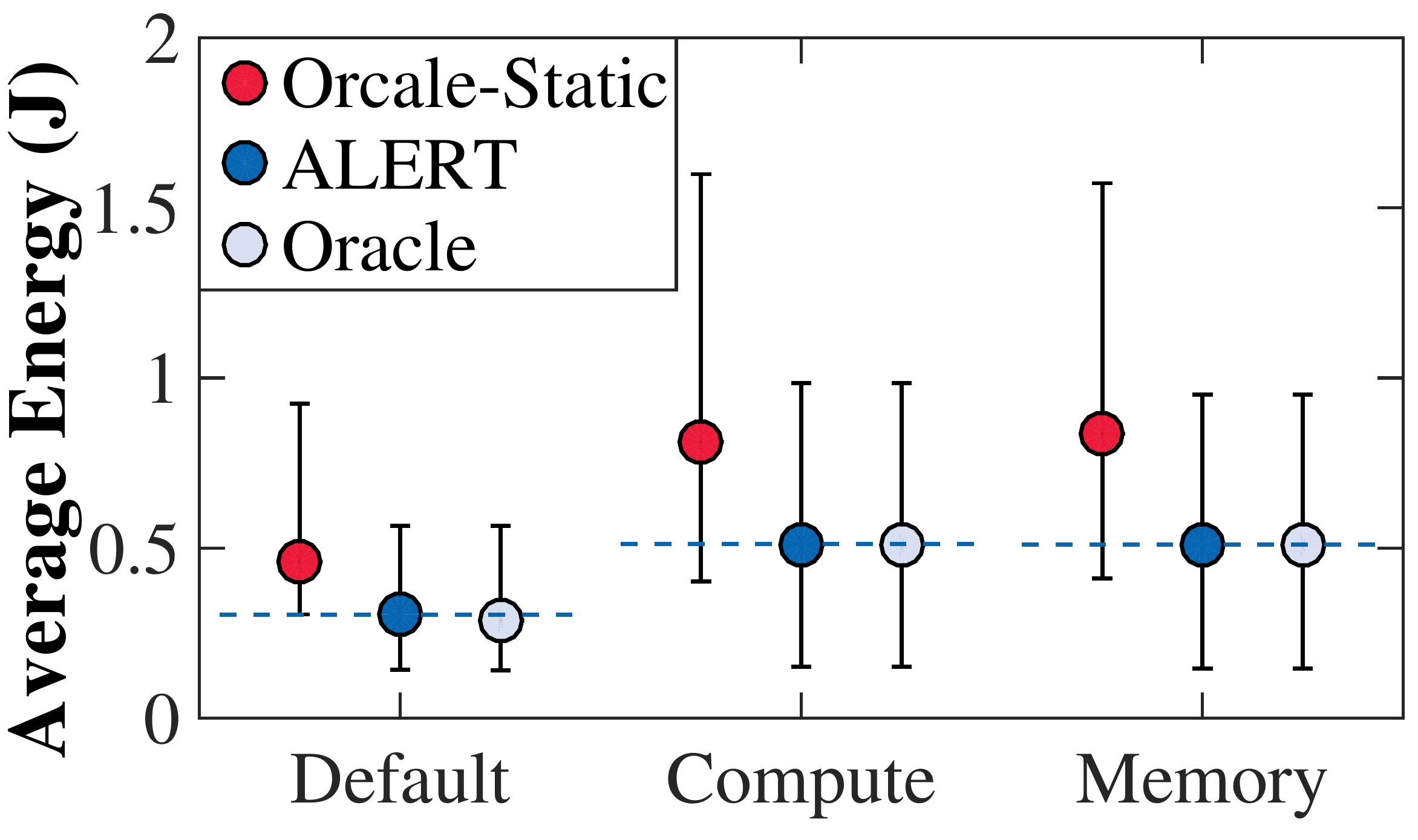}}
  \subfloat[CPU1, Sentence Prediction]{\includegraphics[width=.245\linewidth]{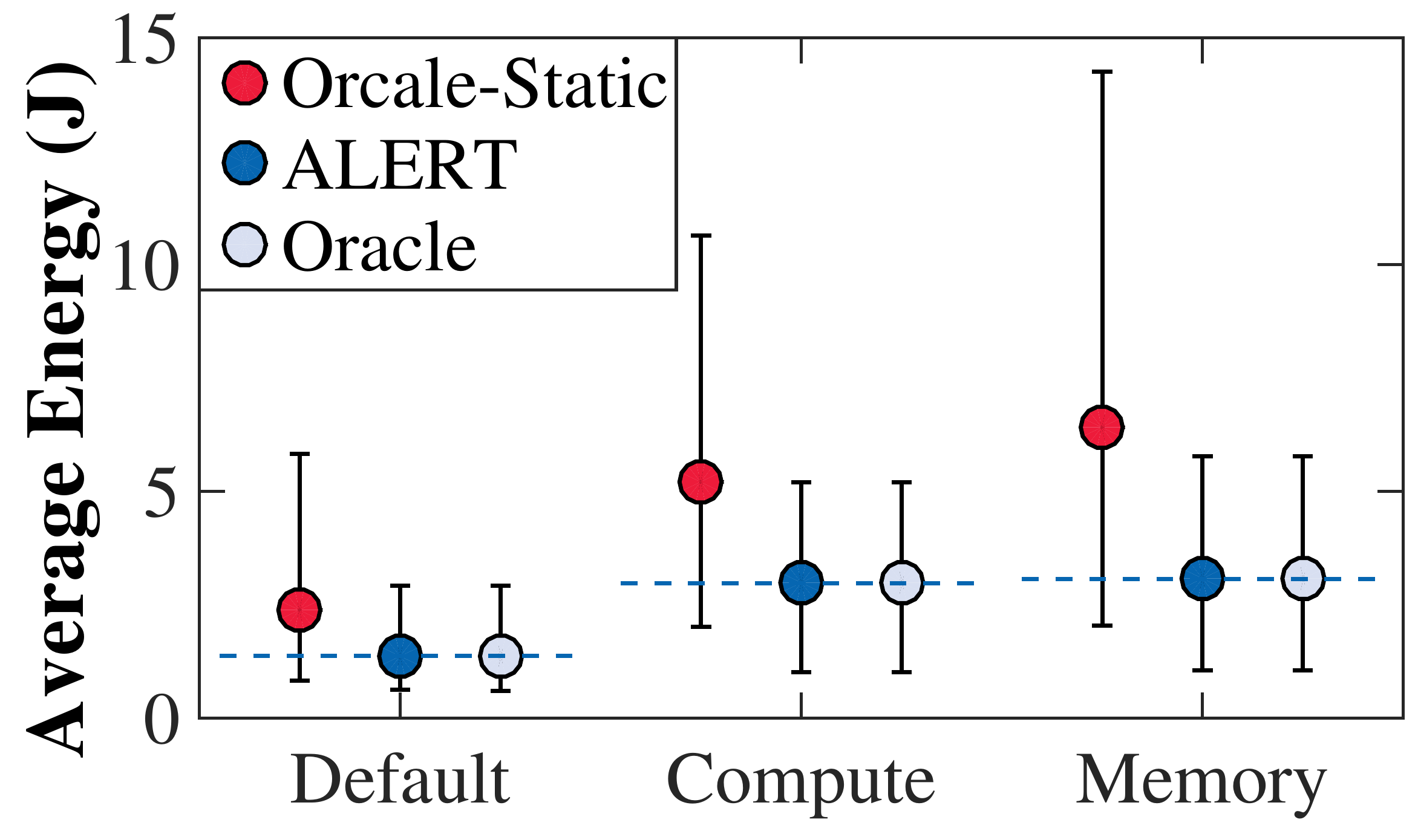}}
  \subfloat[CPU2, Image Classification]{\includegraphics[width=.245\linewidth]{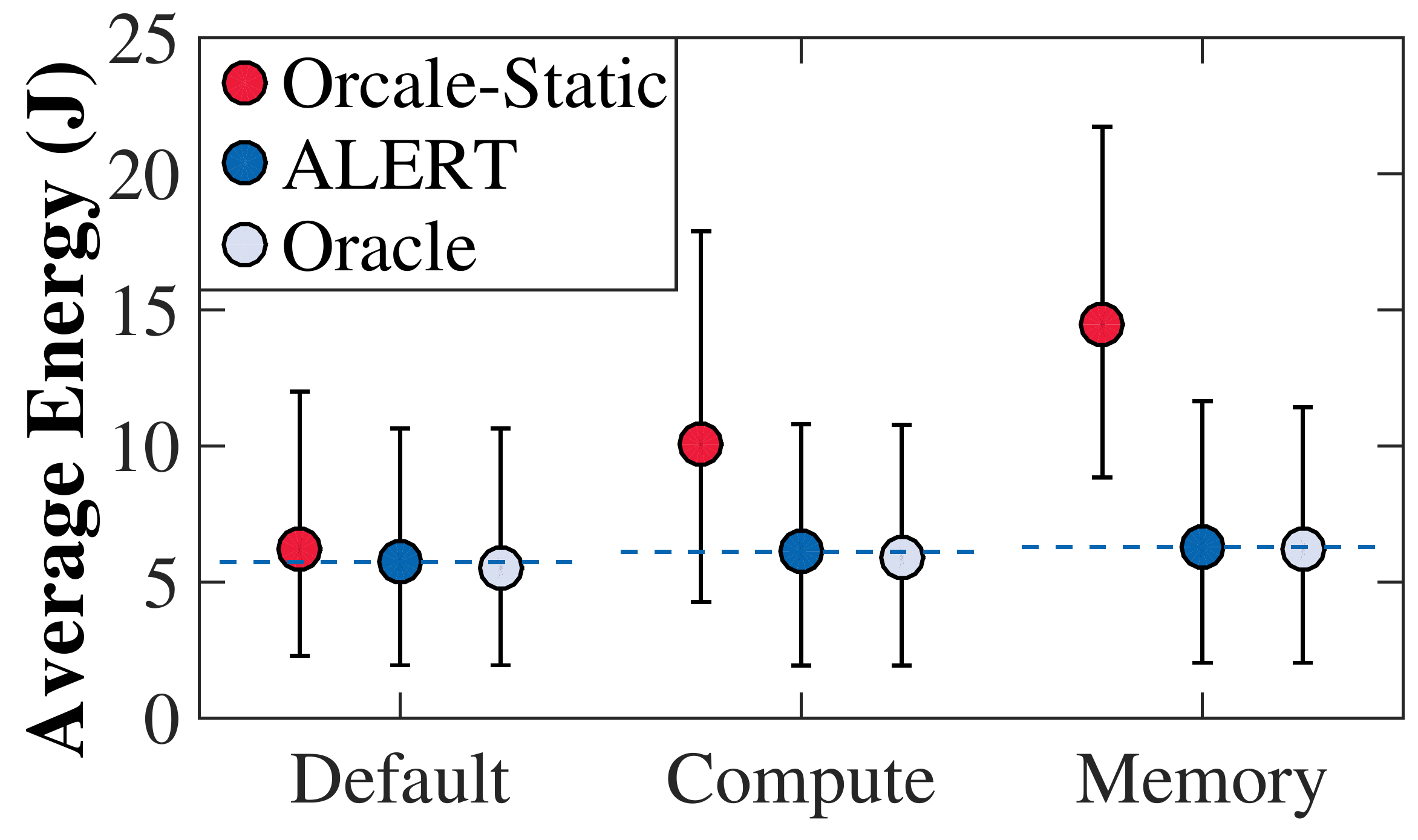}}
  \subfloat[CPU2, Sentence Prediction]{\includegraphics[width=.245\linewidth]{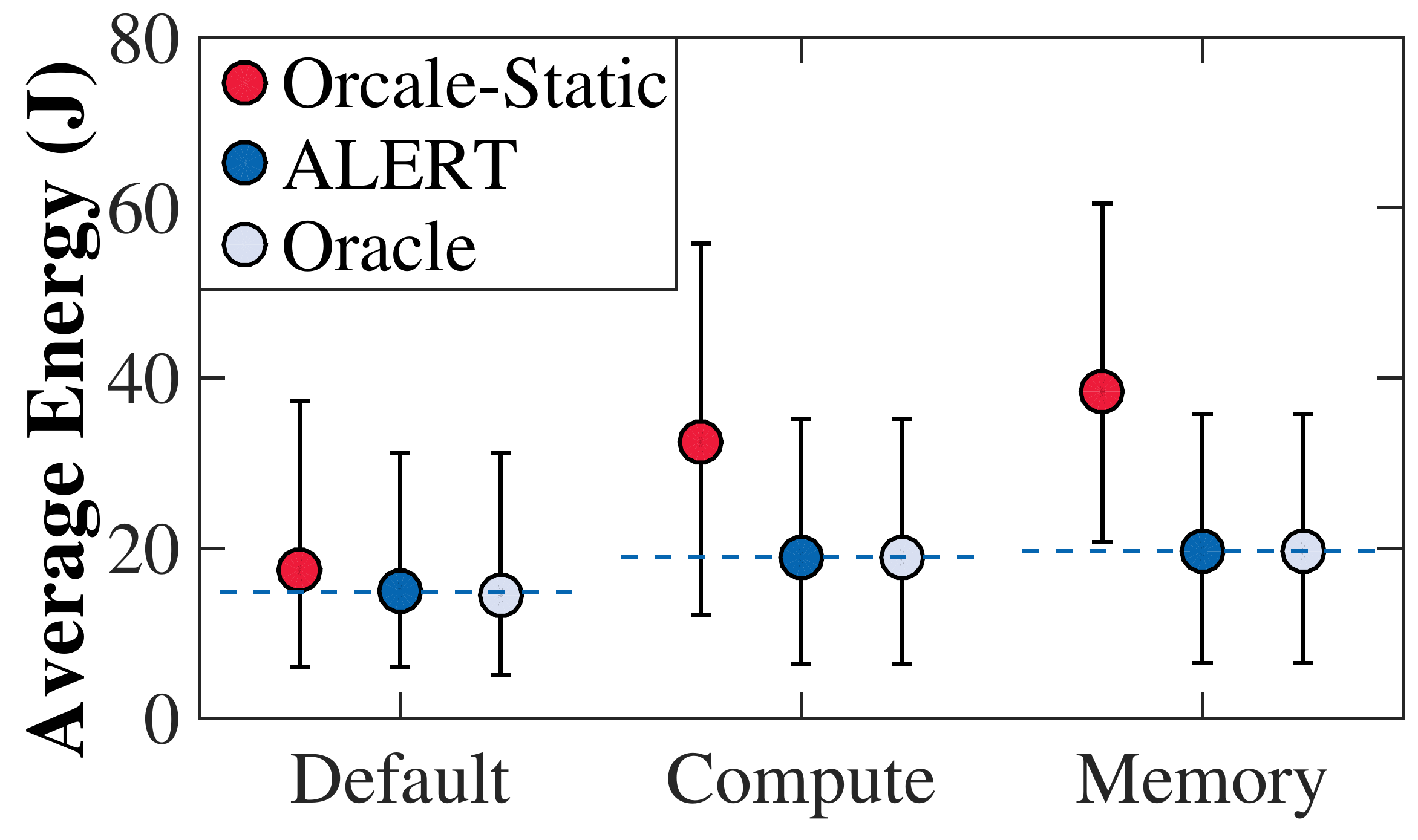}}
  \vspace{-0.11in}
  \caption{ALERT versus Oracle and Oracle$_{\text{Static}}$ on minimize energy task (Lower is better). \textmd{(whisker: whole range; circle: mean)}}
  \label{fig:scheduler1}
\end{figure*}

\textbf{Comparing with Oracles.}
As shown in Table \ref{tbl:result_summary}, \Tool achieves 93-99\% of Oracle's energy and accuracy 
optimization while satisfying constraints.
Oracle$_{\text{static}}$, the baseline in Table \ref{tbl:result_summary}, represents the best one can achieve by selecting
1 DNN model and 1 power setting for all inputs. \Tool  
greatly out-performs Oracle$_{\text{static}}$, reducing
its energy consumption by 3--48\% 
while satisfying accuracy constraints (36\% in harmonic mean) and reducing its error rate by 9-66\% while satisfying energy constraints (54\% in harmonic mean).  

Figure \ref{fig:scheduler1} shows a detailed
comparison for the energy minimization task. The figure shows the range of performance under all requirement settings (i.e., the whiskers). \Tool not only achieves similar
mean energy reduction, its whole range of
 optimization behavior is also similar to Oracle. In comparison, Oracle$_\text{Static}$ not only 
has the worst mean but also the worst tail
performance. Due to space constraints, we omit the figures
for other settings, where similar trends hold.

\Tool has more advantage over Oracle$_\text{static}$ on CPUs than
on GPUs. The CPUs have more empirical variance than the GPU, so they benefit more from dynamic adaptation. The GPU experiences significantly lower dynamic fluctuation so the static oracle makes good predictions.

\Tool satisfies the constraint in 99.9\% of tests for image classification and 98.5\% of those for sentence prediction. For the latter, 
due to the large input variability (NLP1 in Figure \ref{fig:variance}), some
input sentences simply cannot complete by the deadline even with
the fastest DNN. There the Oracle fails, too.

Note that, these Oracle schemes not only have perfect---and hence, impractical---prediction capability, but they also have no overhead. In contrast, \Tool is running on the same machines as the DNN workloads. \emph{All results include \Tool's run-time latency and power overhead.}

\textbf{Comparing with State-of-the-Art.}
For a fair comparison, we focus on \ToolAny, as it uses exactly the same DNN candidate set as "Sys-only", "App-only", and "No-coord".
Across all settings, \ToolAny outperforms the others. 

The System-only solution suffers from not being able to choose different DNNs under different runtime scenarios. As a result, it performs much worse than \ToolAny in satisfying accuracy requirements or optimizing accuracy.
For the former (left side of Table \ref{tbl:result_summary} and Figure \ref{fig:scheduler_summary}), it creates accuracy violations
in 68\% of the settings as shown in Figure \ref{fig:scheduler_summary}; for the latter
(right side of Table \ref{tbl:result_summary} and 
Figure \ref{fig:scheduler_summary}), although capable of satisfying energy constraints, it introduces 34\% more error than \ToolAny.

The Application-only solution that uses an Anytime DNN suffers from not being able to adjust to the energy requirements.
As a result, it consumes 73\% more energy in energy-minimizing tasks (left side of Table \ref{tbl:result_summary} and Figure \ref{fig:scheduler_summary}) and introduces many
energy-budget violations particularly under resource
contention settings (right side of Table \ref{tbl:result_summary} and Figure \ref{fig:scheduler_summary}).

The no-coordination scheme is worse than both System- and Application-only. It violates constraints in both tasks with 69\% more energy and 34\% more error than \ToolAny.  Without coordination, the two levels can work at cross purposes; e.g., the application switches to a faster DNN to save energy while the system makes more power available.  

\begin{table}[!t]
\centering
\footnotesize{
\setlength{\tabcolsep}{0.9mm}
\begin{tabular}{|c|c||c|c|c||c|c|c|}
\hline
\multirow{2}{*}{Plat.} & \multirow{2}{*}{Work.} & ALERT & \begin{tabular}[c]{@{}c@{}}Any\end{tabular} & \begin{tabular}[c]{@{}c@{}}Trad\end{tabular} & ALERT & \begin{tabular}[c]{@{}c@{}}Any\end{tabular} & \begin{tabular}[c]{@{}c@{}}Trad\end{tabular} \\ \cline{3-8} 
                       &                        & \multicolumn{3}{c||}{Minimize Energy Task}                                                                          & \multicolumn{3}{c|}{Minimize Error Task}                                                                           \\ \hline
\multirow{3}{*}{CPU1}  & Idle                   & 0.64  & 0.68                                                & $0.65^{1}$                                           & 0.91  & 0.92                                                & 0.93                                                 \\ \cline{2-8} 
                       & Comp.                    & 0.57  & 0.58                                                & $0.65^{6}$                                           & 0.38  & 0.39                                                & 0.41                                                 \\ \cline{2-8} 
                       & Mem.                   & 0.53  & 0.55                                                & $0.53^{3}$                                           & 0.34  & 0.34                                                & 0.35                                                 \\ \hline
\multirow{3}{*}{CPU2}  & Idle                   & 0.93  & 0.88                                                & $0.95^{1}$                                           & 0.68  & 0.68                                                & 0.69                                                 \\ \cline{2-8} 
                       & Comp.                    & 0.59  & 0.57                                                & $0.60^{4}$                                           & 0.58  & 0.57                                                & 0.59                                                 \\ \cline{2-8} 
                       & Mem.                   & 0.38  & 0.37                                                & $0.40^{8}$                                          & 0.23  & 0.24                                                & 0.32                                                 \\ \hline
\multirow{3}{*}{GPU}   & Idle                   & 0.97  & 0.99                                                & 0.95                                                 & 0.90  & 0.92                                                & 0.89                                                 \\ \cline{2-8} 
                       & Comp.                  & 0.97  & $1.01$                                              & 0.96                                           & 0.89  & 0.92                                                & 0.89                                                 \\ \cline{2-8} 
                       & Mem.                  & 0.96  & 0.97                                                & 0.95                                                 & 0.32  & 0.34                                                & 0.32                                                 \\ \hline
\rowcolor[rgb]{0.749,0.749,0.749} \multicolumn{2}{|c||}{Harmonic mean}   & 0.66            & 0.66                                                 & $0.67^{3}$                                            & \multicolumn{1}{c|}{0.47} & 0.48                                                & 0.50                                                  \\
\hline
\end{tabular}
}
\caption{ALERT normalized average energy consumption and error rate to \emph{Oracle}$_\text{Static}$ @ Sparse ResNet (Smaller is better)}
\vspace{-0.15in}
\label{tbl:result_dnn}
\end{table}

\subsection{Detailed Results and Sensitivity}
\label{sec:eval_details}

\textbf{Different DNN candidate sets.}
Table \ref{tbl:result_dnn} compares the performance 
of \Tool working with an Anytime DNN (Any), a set of traditional
DNN models (Trad), and both.
At a high level, \Tool works well with all three DNN sets. 
Under close comparison,  
\ToolTrad violates more accuracy constraints than the others, particularly under resource contention on CPUs, because a traditional DNN has a much larger
accuracy drop than an anytime DNN when missing a latency deadline.
Consequently,
when the system variation is large, \ToolTrad selects a faster DNN to meet latency and thus may not meet accuracy goals.
Of course, \ToolAny is not always the best. 
As discussed in Section \ref{sec:design_any}, Anytime DNNs sometimes have lower accuracy then a traditional DNN with similar execution time. This difference leads to the slightly better results for \Tool over \ToolAny.

\begin{figure}
  \centering
  \includegraphics[width=.95\linewidth]{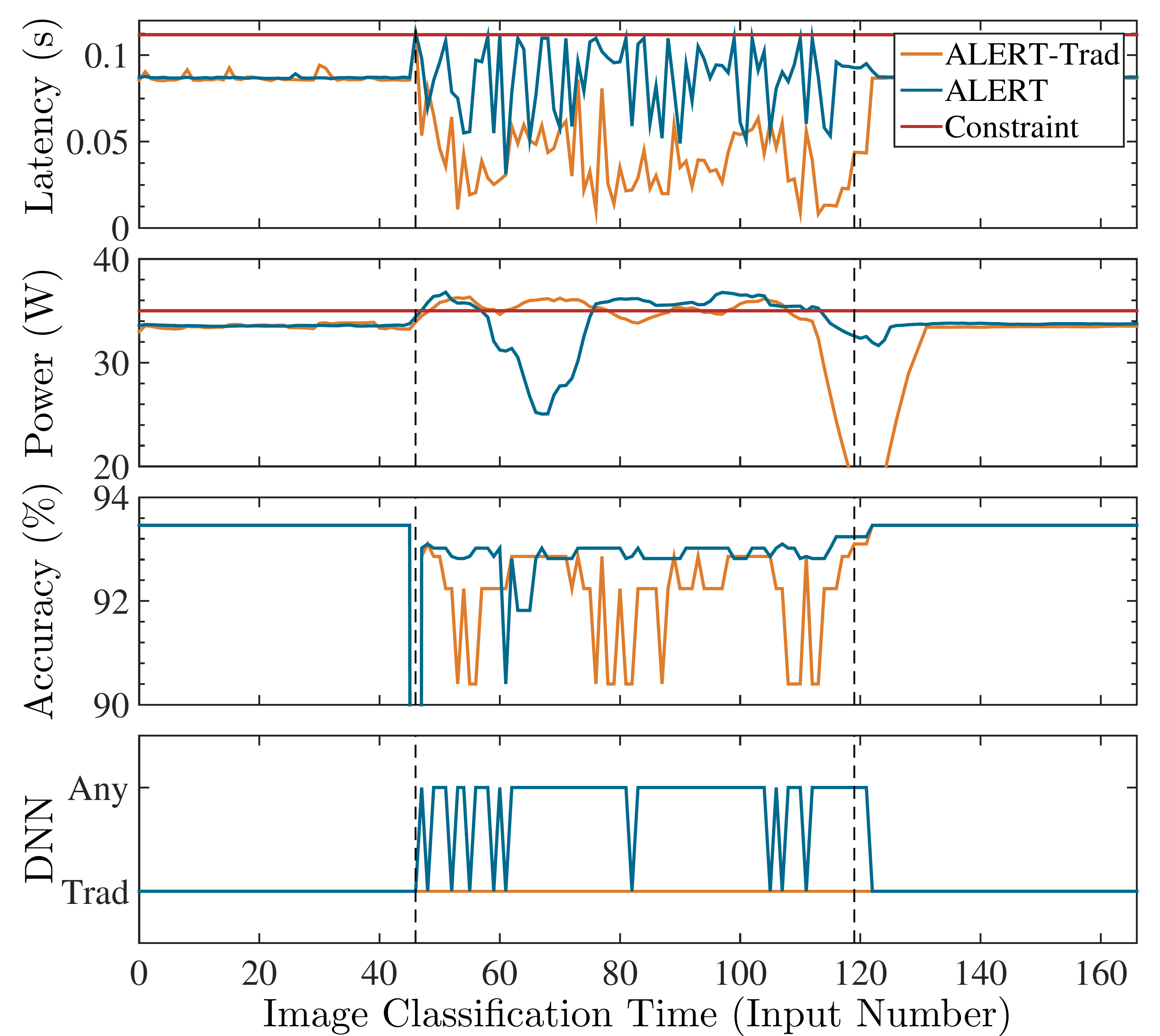}
\vspace{-0.1in}
  \caption{Minimize error rates w/ latency, energy constraints @ CPU1. (\Tool in blue; ALERT$_\text{Trad}$ in orange; constraints in red. Memory contention occurs from about input 46 to 119; Deadline: 1.25$\times$ mean latency of largest Anytime DNN in Default; power limit: 35W.)}
\vspace{-0.15in}
  \label{fig:co-locate}
\end{figure}

Figure \ref{fig:co-locate} visualizes the different dynamic behavior of \Tool (blue curve) and ALERT$_{\text{Trad}}$ (orange curve) when the environment changes from Default to Memory-intensive and back. 
At the beginning, due to a loose latency constraint, \Tool and  ALERT$_{\text{Trad}}$ both
select the biggest traditional DNN, which provides the highest accuracy within the energy budget.
When the memory contention suddenly starts, this DNN choice leads to a deadline miss and an energy-budget violation (as the idle period disappeared), which causes an accuracy dip.  Fortunately, both 
quickly detect this problem and sense the high variability in the expected latency.
\Tool switches to use an anytime DNN and a lower power cap. This switch is  effective:
although the environment is still unstable, the inference accuracy remains high, with slight ups and downs depending on which anytime output finished before the deadline.  
Only able to choose from traditional DNNs, ALERT$_{\text{Trad}}$ conservatively switches to much simpler and hence lower-accuracy
DNNs to avoid deadline misses. This switch does eliminate deadline misses under the highly dynamic environment, but many of the conservatively chosen DNNs finish before the deadline (see the Latency panel), wasting the opportunity to produce more accurate results and causing ALERT$_{\text{Trad}}$ to have a lower accuracy 
than \Tool. When the system quiesces, both schemes quickly shift back to the highest-accuracy, traditional DNN. 

Overall, these results demonstrate how \Tool 
always makes use of the full potential of the DNN candidate set to optimize performance and satisfy constraints.

\textbf{\Tool probabilistic design.}
A key feature of \Tool is its use of not just mean estimations, but also their variance.  To evaluate the impact of this design, we compare \Tool to an alternative design \ToolAlter, which only uses the estimated mean to select configurations.

\begin{figure}[t]
  \centering
    \vspace{-0.2in}
  \subfloat[Default Contention]{\includegraphics[width=.49\linewidth]{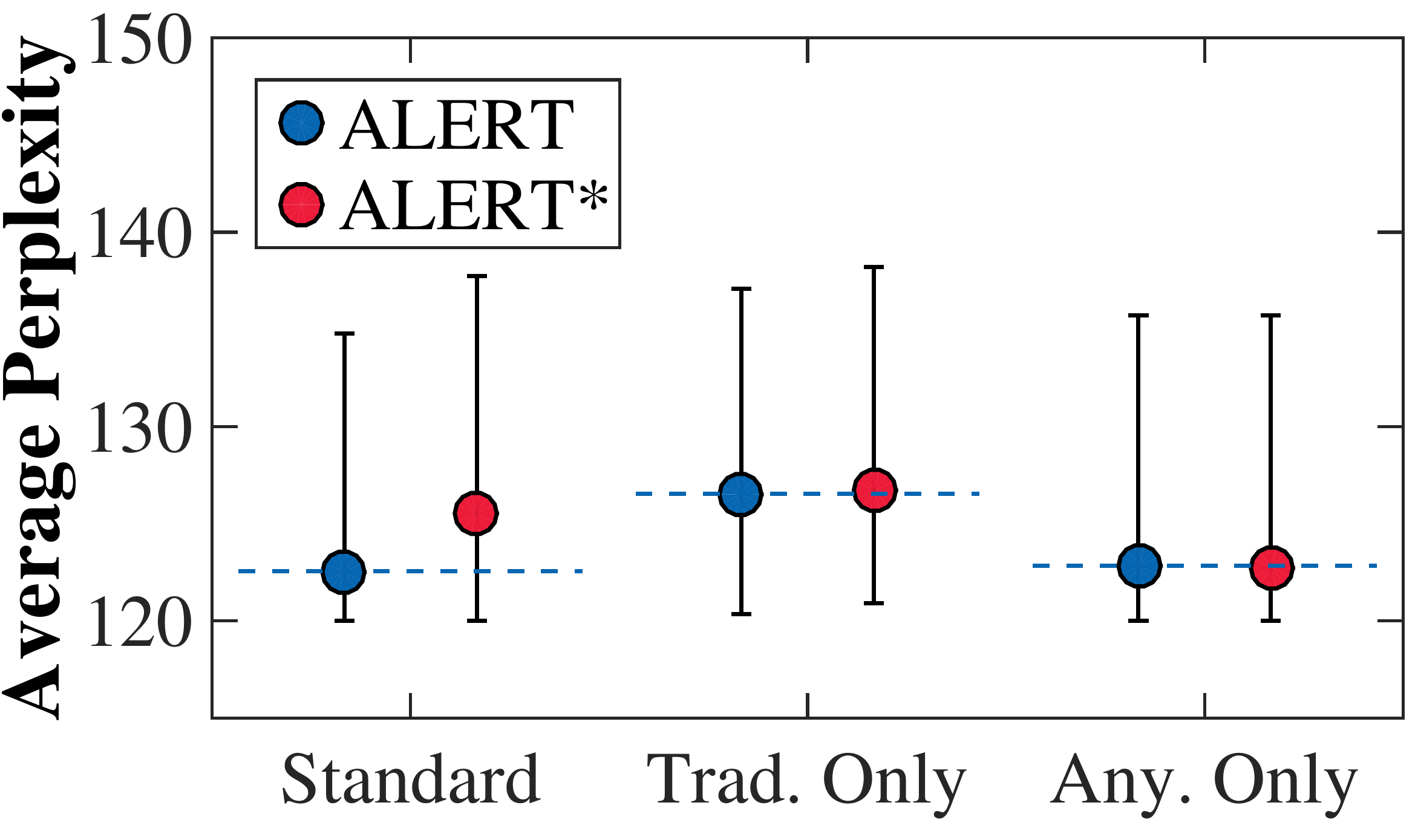}}
  \subfloat[Memory Contention]{\includegraphics[width=.49\linewidth]{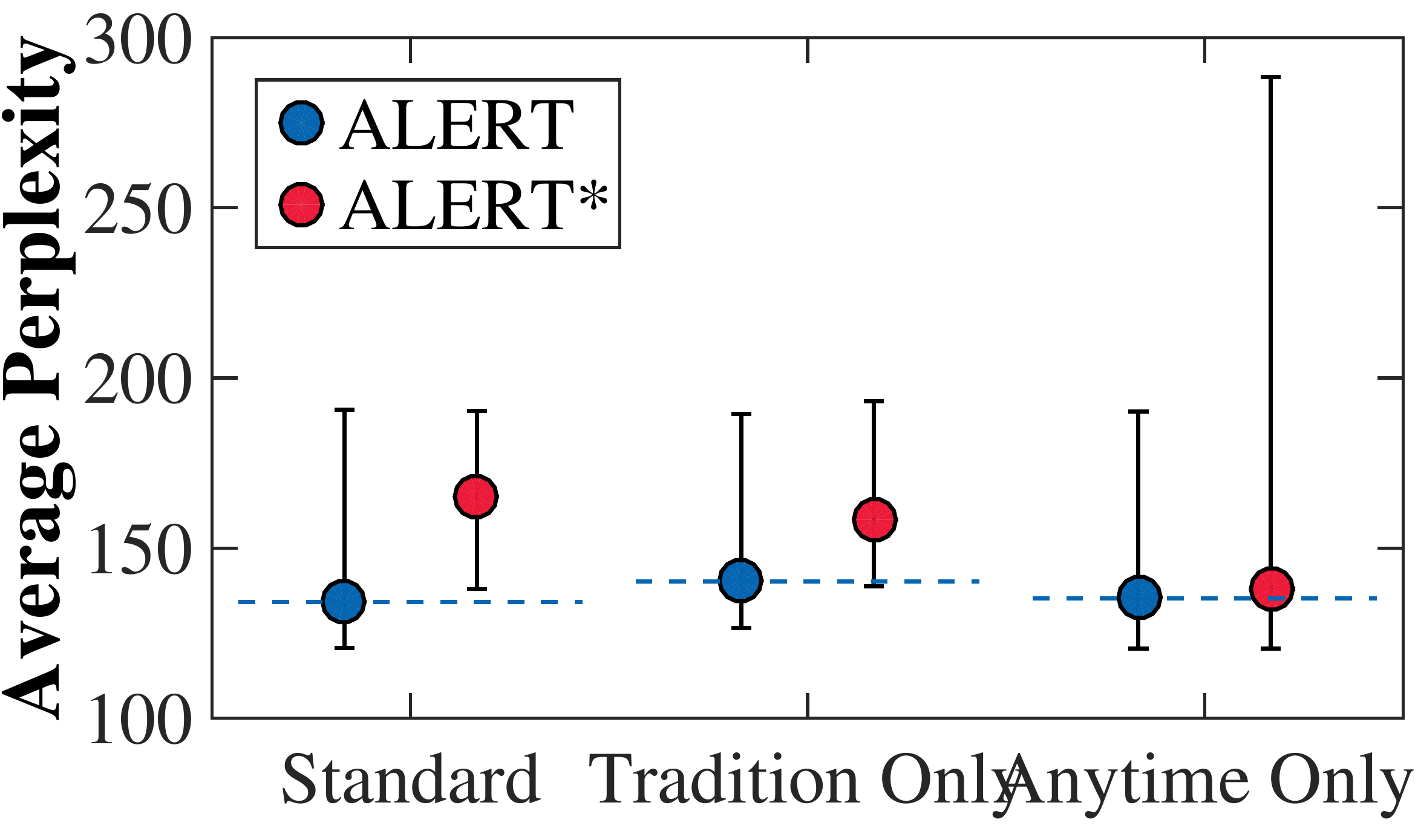}}
  \vspace{-0.11in}
  \caption{Minimize error for sentence prediction@ CPU1 (Lower is better). \textmd{(whisker: whole range; circle: mean)}}
\vspace{-0.05in}
  \label{fig:alert-alter}
\end{figure}

Figure \ref{fig:alert-alter} shows the performance of \Tool and \ToolAlter in the minimize error task for sentence prediction. As we can see, \Tool (blue circles) always
performs better than \ToolAlter. Its advantage is the biggest when the DNN candidate sets include both traditional and Anytime DNNs, which is the ``Standard'' in Figure \ref{fig:alert-alter}. The reason is that
traditional DNNs and Anytime DNN have different 
accuracy/latency curves, Eq. \ref{eq:expected_acc}
for the former and Eq. \ref{eq:expected_acc_any} for the latter. \ToolAlter is much worse than \Tool in distinguishing these
two by simply using the mean of estimated latency
to predict accuracy. \Tool's advantage is also reflected under memory contention with traditional DNN candidates. Since \Tool's estimation better
captures dynamic system variation, it clearly
outperforms \ToolAlter there.
Overall, these results show \Tool's probabilistic design is effective.

\begin{figure}[t]
    \centering
    \includegraphics[width=0.99\linewidth]{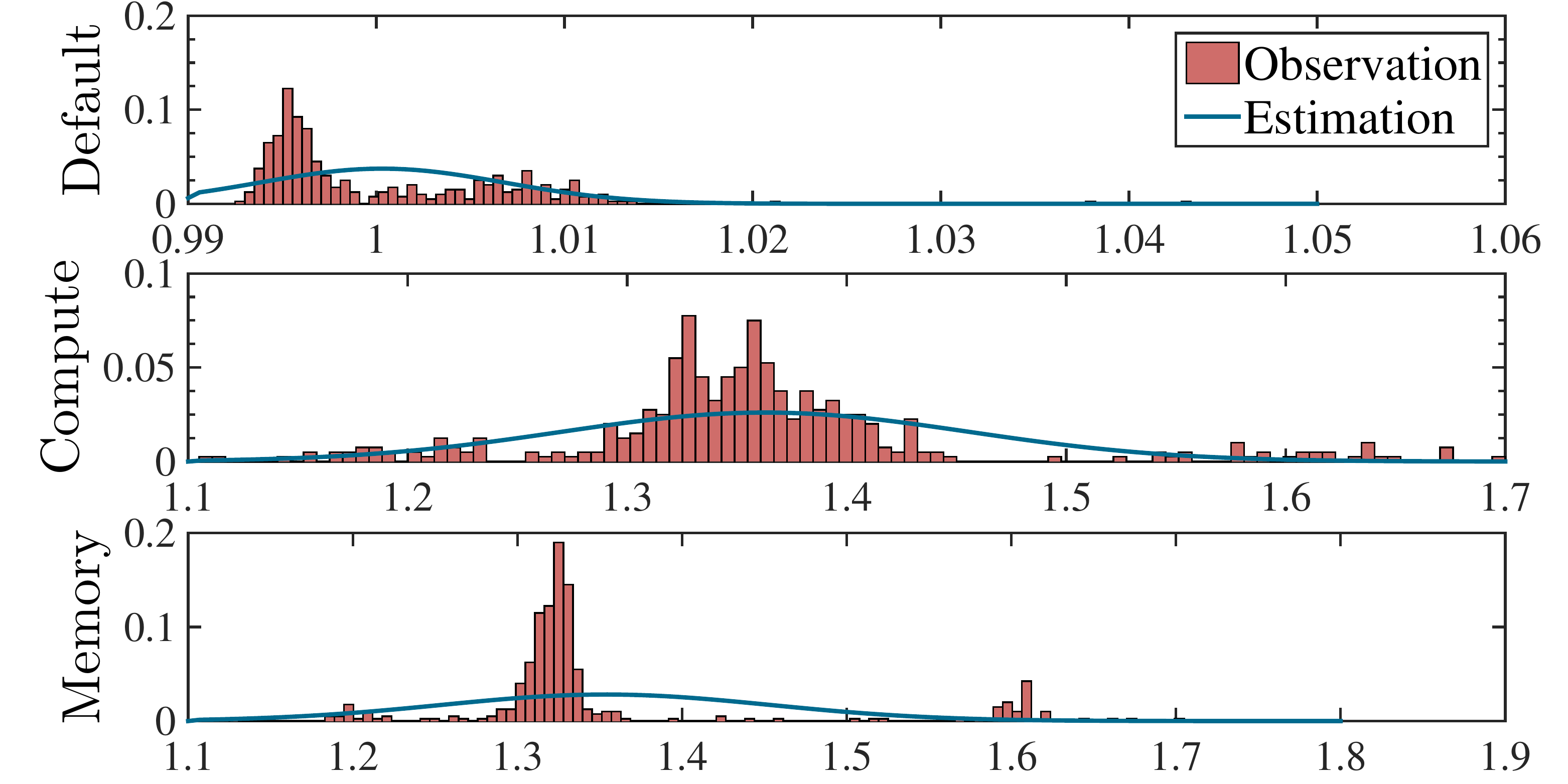}
    \vspace{-0.11in}
    \caption{Distribution of $\xi$ for image class. on CPU1. }
    \label{fig:distribution}
    \vspace{-0.05in}
\end{figure}

\textbf{Sensitivity to latency distribution.}
\Tool assumes a Gaussian distribution. 
However, \Tool is still robust for other distributions, as explained in Section \ref{sec:design_limit}.
As shown in Figure \ref{fig:distribution},
the observed $\xi$s (red bars) are 
indeed not a perfect fit for Gaussian distribution (blue lines) in all scenarios, which confirms \Tool's robustness.

\section{Related work}
\label{sec:related}

Past resource management systems have used machine learning \cite{ansel2012siblingrivalry, lee2008efficiency, petrica2013flicker, ponomarev2001reducing, sridharan2013holistic} or control theory \cite{kalman2009,kalman2014, caloree, hoffmann2015, santriaji2016grape, zhou2016cash,poet} to make dynamic decisions and adapt to changing environments or application needs.  Some also use Kalman filter because it has optimal error properties \cite{poet,kalman2009,kalman2014,caloree}.  There are two major differences between them and \Tool:  1) prior approaches use the Kalman filter to estimate physical quantities such as CPU utilization \cite{kalman2014} or job latency \cite{poet}, while \Tool estimates a \emph{virtual} quantity that is then used to update a large number of latency estimates. 2) while variance is naturally computed as part of the filter, \Tool actually uses it, in addition to the mean, to help produce estimates that better account for environment variability.

Past work designed resource managers explicitly to coordinate approximate applications with system resource usage \cite{hoffmann2015,coadapt,lab_abstraction,farrell2016}. Although related,
they manage applications
{\it separately} from system resources, which is
fundamentally different from \Tool's holistic design. When an environmental change occurs, prior approaches 
first adjust the application and then the system serially (or vice versa) so that the change's effects on each can be established independently \cite{coadapt,hoffmann2015}. That is, coordination is established by forcing one level to lag behind the other. In practice this design forces each level to keep its own independent model and 
delays response to environmental changes.  
In contrast, \Tool's global slowdown factor allows it to easily model and update
prediction about all application and system configurations simultaneously, leading to very fast response times, like the single input delay demonstrated in Figure \ref{fig:co-locate}. 


Much work accelerates DNNs 
through hardware \cite{chen2014diannao, chen2014dadiannao, hauswald2015djinn, du2015shidiannao, liu2015pudiannao, rhu2016vdnn, albericio2016cnvlutin, eyeriss, han2016eie, judd2016stripes, hill2017deftnn,
gao2016draf, mahajan2016tabla, sharma2016high, ovtcharovaccelerating,
vanhoucke2011improving, jain2018architectural},
compiler \cite{chen2018tvm, tensorRT}, 
system \cite{hauswald2015sirius, lin2018architectural}, or
design support \cite{venkataramani2014axnn, han2015compression, judd2016proteus, hashemi2017understanding, tann2017hardware, jain2018compensated, sim2018dps, han2015compression}.
They essentially shift and extend the tradeoff space,
but do not provide policies for meeting user needs or for navigating tradeoffs dynamically, and
hence are orthogonal to \Tool.

Some research supports hard real-time guarantees for DNNs \cite{S3DNN}, providing 100\% timing guarantees while assuming that the DNN model gives the desired accuracy, the environment is completely predictable, and  energy consumption is not a concern.  \Tool provides slightly weaker timing guarantees, but manages accuracy and power goals. \Tool also provides more flexibility to adapt to unpredictable environments. Hard real-time systems would fail in the co-located scenario unless they explicitly account for all possible co-located applications at design time.

\section{Conclusion}
\label{sec:conclusion}

This paper demonstrates the challenges behind the important problem of ensuring timely, accurate, and energy efficient neural network inference with dynamic input, contention, and
requirement variation. 
\Tool achieves these goals through dynamic and coordinated DNN model selection and power management based on feedback control.
We evaluate \Tool with a variety of workloads and DNN models and achieve high performance and energy efficiency.

\section*{Acknowledgement}
We thank the anonymous reviewers for their helpful feedback and Ken Birman for shepherding this paper.
This research is supported by NSF (grants CNS-1956180, CNS-1764039, CCF-1837120, CNS-1764039, CNS-1563956, CNS-1514256, IIS-1546543, CNS-1823032, CCF-1439156), ARO (grant W911NF1920321), DOE (grant DESC0014195 0003), DARPA (grant FA8750-16-2-0004) and the CERES Center for Unstoppable Computing. Additional support comes from the DARPA BRASS program and a DOE Early Career award.

%
\bibliographystyle{plain}
\bibliography{citations} 
\balance

\end{document}